\documentstyle[eqsecnum,psfig,aps,prb,multicol]{revtex}
\begin{document} 
\draft

\title{Nodal Liquid Theory of the Pseudo-Gap Phase of High-${T_c}$
Superconductors}
\author{Leon Balents,
Matthew P. A. Fisher, and Chetan Nayak} 
\address{Institute for Theoretical Physics, University of 
California,
             Santa Barbara, CA 93106-4030}
\date{\today} 
\maketitle

\begin{abstract}

We introduce and study the {\sl nodal liquid}, a novel zero-temperature 
quantum phase obtained by quantum-disordering a d-wave superconductor.  
It has numerous remarkable properties which lead us to
suggest it as an explanation of the pseudo-gap state in 
underdoped high-temperature superconductors.  In the absence of 
impurities, these include power-law 
magnetic order, a $T$-linear spin susceptibility, non-trivial thermal 
conductivity, and two- and one-particle charge gaps, the latter evidenced, 
e.g. in transport and electron photoemission (which exhibits pronounced 
fourfold anisotropy inherited from the d-wave quasiparticles).  
We use a $2+1$-dimensional duality transformation
to derive an effective field theory for this phase.
The theory is comprised of gapless {\sl neutral}
Dirac particles living at the former d-wave nodes, {\sl weakly
coupled} to the fluctuating gauge field of a dual
Ginzburg-Landau theory. The nodal liquid interpolates
naturally between the d-wave superconductor and the 
insulating antiferromagnet, and our effective field
theory is powerful enough to permit a detailed analysis
of a panoply of interesting phenomena, including
charge ordering, antiferromagnetism, and d-wave superconductivity.  
We also discuss the zero-temperature quantum
phase transitions which separate the nodal liquid
from various ordered phases.

\end{abstract}
\vskip 0.5 in
\begin{multicols}{2}

\section{Introduction}

The discovery of the cuprate high-temperature
superconductors in 1986\cite{Bednorz}\ was a watershed in the
recent history of condensed matter physics,
an event which stimulated intense
experimental and theoretical activity.
As sample quality and experimental precision have
advanced, these materials' rich phase diagram and
phenomenology have come into focus.\cite{Maple}\
However, many theoretical efforts
have not reached fruition because of a
serious obstacle, namely, that these materials are apparently
in a strongly-coupled, non-perturbative regime.
To put it more bluntly: there is no obvious
small parameter which facilitates an expansion about
a tractable model.
In this paper, we promulgate the existence
of a {\sl weakly-coupled} `dual' description of a particularly exotic
region of the phase diagram. This description paves
the way for controlled calculations of experimentally
measurable quantities.

Before plunging into our exegesis of this
dual description and its consequences, let us briefly review
the phenomena which we wish to explain.
The phase diagram as a function of temperature $T$ and 
doping $x$ (the nature of the doping varies from material to material, 
but is generally believed to be proportional to the effective hole 
concentration in each $CuO_2$ layer) is indicated
schematically in Fig.~1. Best understood and in many 
cases very well characterized are the undoped ($x=0$) materials, which 
are insulating antiferromagnets (AFs) below the N\'eel temperature 
$T_N$.  At moderate dopings ($0.1\lesssim x \lesssim 0.3$) and low 
temperatures ($T<T_{c}(x)$), superconductivity occurs. 
For many phenomenological purposes the superconducting
phase is adequately described by the same 
Ginzburg-Landau and London theory standard for conventional 
superconductors (SCs).\cite{Tinkham}\  One important distinction between the 
high-T$_c$ and conventional SCs, suspected for many years and now 
generally accepted after a number of beautiful and compelling 
experiments,\cite{Wollman,Kirtley}\ is their $d_{x^2-y^2}$ ({\sl d-wave}) 
pairing symmetry.  This pairing symmetry is a crucial
ingredient in a zero-temperature quantum 
description. In particular, d-wave symmetry leads to
gapless quasiparticles residing at the four nodes of the 
pair wavefunction.

\begin{figure}
\psfig{figure=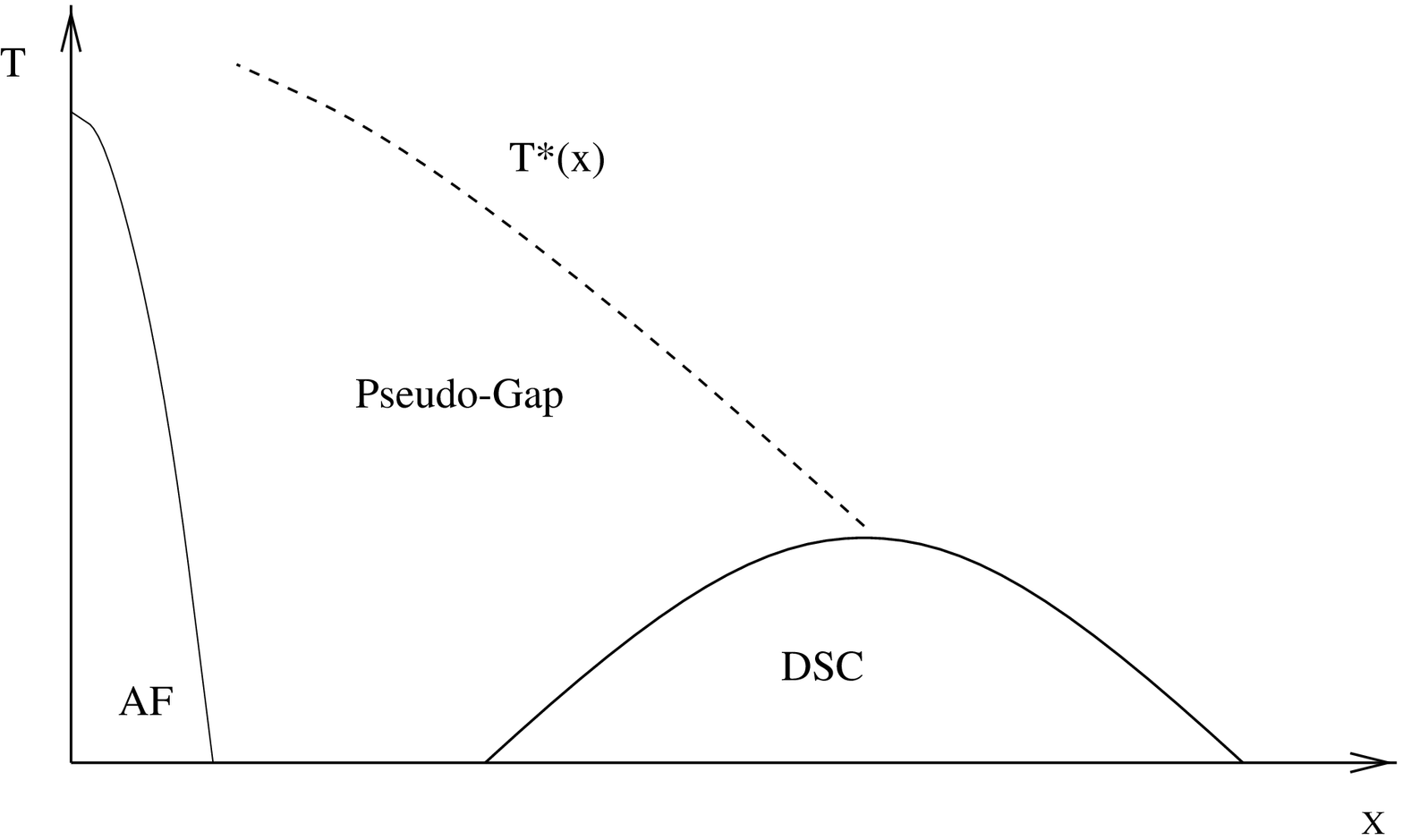,height=2.0in,angle=-90}
\vskip 0.5cm
{Fig.~1: Schematic phase diagram of a high-temperature superconductor
  as a function of doping $x$ and temperature $T$.}
\end{figure}

Recent studies have revealed puzzling behavior in the {\sl underdoped}
region between the AF and SC. Below the high-temperature dotted line
($T^*(x)$) in Fig.~1, angle-resolved photoemission
(ARPES),\cite{Loeser,Ding}\ transport,\cite{Batlogg}\ 
NMR,\cite{Warren,Takigawa}\ optical conductivity,\cite{Homes,Puchkov}\ 
and other measurements\cite{Maple}\ indicate a dramatic reduction of
low-energy (single-)electronic and spin degrees of freedom.
Furthermore, the ability of ARPES measurements to resolve wavevector
dependence exposes an angular variation similar to that of d-wave
quasiparticles in the SCing phase\cite{Loeser,Ding}.  This portion of the
phase diagram is commonly called the {\sl pseudo-gap} region.  The
ultimate nature of the corresponding underlying quantum ground state
is an intriguing theoretical puzzle, and a principal subject of this
paper.

To proceed, we look to the experiments for guidance.
They indicate three types of excitations which are
important below the dotted line in Fig.~1: the ordering
fields related to antiferromagnetism and superconductivity,
{\it and} d-wave quasiparticles near the four nodes.
Conspicuously absent from this 
list are electrons and holes at an ordinary Fermi-surface.  The 
physics of this omission is that pairing occurs (due to unspecified strong 
interaction physics) at the high energy $T^*(x)$.  Given these 
ingredients, one natural theoretical strategy is to attempt to 
approach the pseudo-gap state by increasing $x$ from the AF at 
half-filling.  Many researchers have already attempted this approach, 
but it remains inconclusive.  

We, instead, tackle the pseudo-gap state from the
right, literally. To do this, we must
contemplate quantum-disordering the d-wave superconductor.
For simplicity, we will assume for the moment a purely two-dimensional 
model of a single $CuO_{2}$ plane.  We 
imagine that pairing establishes a local superconducting d-wave order
parameter $\Delta(\vec{x},t) = |\Delta| e^{i\varphi}$, where ${\vec{x}}$ is 
the two-dimensional coordinate and $t$ is time.
The experimental properties of
the pseudo-gap state can be interpreted as
an indication that $|\Delta|$ is large in this region, 
so that quantum fluctuations of the phase of the order-parameter
$\varphi$ must be responsible for the lack
of off-diagonal long-range-order ($\langle 
\Delta \rangle = 0$), even as $T \rightarrow 0$.  The important long-distance 
dephasing is 
accomplished by {\sl vortex} loops and lines, around which $\varphi$ 
winds by $\pm 2\pi$.  To destroy the long-range correlations in 
$\Delta$, we must unbind vortex loops of arbitrarily large size, just 
as vortex-antivortex pairs unbind above the Kosterlitz-Thouless 
transition temperature in a two-dimensional superfluid.  To implement 
this unbinding, it is extremely helpful to use the $2+1$-dimensional {\sl 
duality}\cite{fDuality,Duality}\ relating an XY-model for $\Delta$ and a  
Ginzburg-Landau model with complex field $\Phi$ (``disorder parameter'') 
interacting  with a gauge field ${\bf a}$.  The duality interchanges Cooper 
pairs and vortices, so that the desired quantum disordered d-wave 
state is the ordered (condensed) phase, $\langle \Phi\rangle \neq 0$, 
of the Ginzburg-Landau theory. The imaginary-time
effective action for this dual theory is nothing
but the Ginzburg-Landau free energy functional
for a three-dimensional superconductor at finite-temperature.

The fate of the d-wave quasiparticles in this construction requires 
particular care, since these are strongly coupled to the fluctuating 
pair field.  In Sec.~\ref{sec:duality}, we show 
how the quasiparticles can be treated by extracting the $U(1)$ phase 
from the bare electron operators.  Once the phase is extracted, one is 
left with a set of gauge-invariant fermion operators which create 
electrically neutral (but spin-ful) quanta we call {\sl nodons}.
In the dual variables, the 
fundamental Lagrangian of our model is
\begin{eqnarray}
{\cal L} &=& \psi_1^\dagger [ i \partial_t - v_F \tau^z
  i\partial_x - v_{\scriptscriptstyle\Delta} 
  \tau^x i\partial_y ] \psi_1  + (1 \leftrightarrow 2,x \leftrightarrow y)\cr
& & + {\kappa_\mu \over 2} |(\partial_\mu - ia_\mu)\Phi|^2 - V_{\Phi}
  (|\Phi|)\cr
& &  + a_\mu \epsilon_{\mu \nu \lambda} 
\partial_\nu (A_\lambda - \kappa_\lambda^{-1} J_\lambda)
+ {1 \over {2\kappa_0}} (e_j^2 - b^2)\cr
& & + 2\lambda J_0 + {{\cal L}_{N}}  .
\label{eq:lagrangian}
\end{eqnarray}
In the sections which follow, we will elucidate the physics
of this Lagrangian in some detail, so we restrict
ourselves, in this introduction, to a whirlwind tour.
The fields ${\psi^\dagger_j},{\psi_j}$, $j=1,2$ are the nodon
creation and annihilation operators at the two antipodal
pairs of nodes. $J_\lambda$ is a bilinear in
the nodon operators which has an interpretation in
the d-wave superconducting phase as
the quasiparticle electrical 3-current. $\Phi$ is the
complex scalar field representing the
vortices, and $a_\mu$ is the gauge field which
is dual to the phase of the superconducting
order parameter. The term proportional to $\lambda$
describes the effects of particle/hole asymmetries, and
the ${\cal L}_{N}$ term describes the coupling of the 
nodons to antiferromagnetism, which we will return to presently.  

A remarkable result of calculations with Eq.~\ref{eq:lagrangian}\ is
that gapless nodons survive the quantum disordering of the SC!  The
nodons are like the smile of the Cheshire cat: the $d_{x^2-y^2}$ order
parameter is gone, but the nodes remain.  The consequent {\sl Nodal
  Liquid} (NL) described by Eq.~\ref{eq:lagrangian}\ is a distinct and
novel zero-temperature quantum phase with a number of fascinating
properties.  For simplicity, consider first a hypothetical NL phase at
half-filling in the absence of impurities.  The possibility of AF
ordering will be included later via ${\cal L}_N$.  We hypothesize that
antiferromagnetism might be avoided and the NL achieved in a
half-filled square lattice model by adjusting an attractive
nearest-neighbor interaction and second-neighbor electron hopping
amplitude.  The NL is a nominally insulating state, with non-zero gaps
$\Delta_1$ and $\Delta_2$ for adding both individual electrons/holes
and Cooper pairs, respectively.  Gapless nodons with anisotropic
ballistic dispersion ($\omega \sim k$), however, persist, and can
carry both spin and current.  With particle/hole symmetry ($\lambda=0$
in Eq.~\ref{eq:lagrangian}), we expect power law ($\sim 1/|x|^{4}$)
spin correlations at $(0,0)$, $(\pi,\pi)$, $(\pi,0)$, and
symmetry-related points in the Brillouin zone.  Scaling arguments lead
us to expect a weak dissipative dynamic contribution to the
conductivity which, in the presence of particle/hole symmetry, would
vary as ${\rm Re}\, \sigma(\omega,T=0) \sim \omega^{6}$.  Nodons also
contribute a quadratic specific heat $C_{\rm nodon} \sim a T^2$.
Despite the similarity of the nodons to d-wave quasiparticles, the
single-particle spectral function is predicted to show a gap at the
Fermi energy in the NL.  We expect, however, this gap to be strongly
angle-dependent: of order the pairing scale $T^*$ away from $(\pm
\pi/2,\pm \pi/2)$ and reduced to $\Delta_1 \ll T^*$ at these special
points.  A particle/hole asymmetric NL should exhibit similar
behavior, but with singularities shifted from $(\pm \pi/2,\pm \pi/2)$
in momentum space, and an even smaller contribution to the
low-frequency dissipative conductivity which, in the absence of
umklapp scattering and impurities must come from interactions with
phonons.

Consider next doping the NL.  Naively, this can be modeled via an
increase in the in-plane hole concentration, although the actual
transfer of charge to the $CuO_2$ layers may be not be complete.  In
the grand canonical ensemble, charge is added by increasing $\mu =
A_0/2$.  In the dual theory (see Eq.~\ref{eq:lagrangian}), this chemical
potential acts like an external magnetic field $h=2\mu$ in the
Ginzburg-Landau theory (but, unlike a magnetic field, it of course
does not break time-reversal invariance).  For small $\mu$, the system
remains in the Meissner phase, and no dual flux penetrates -- i.e. no
charge is added to the system within the charge gap. Following the
analogy with Ginzburg-Landau theory,\cite{Tinkham}\ we expect that the nature of
doping depends upon the Abrikosov parameter $\kappa_v =
\lambda_v/\xi_v$, where $\lambda_v$ and $\xi_v$ are effective dual
penetration and coherence lengths, respectively.  For $\kappa_v
\gtrsim 1/\sqrt{2}$, {\sl type II} doping occurs, and the ``field''
penetrates first for $\mu>\mu_{c1}$ in a dual flux lattice.  Dual flux
tubes are in fact Cooper pairs, so this is a paired Wigner crystal
(PWC) state. We show in Sec.~\ref{sec:NL}\ that gapless nodons survive
the doping and coexist with the PWC.  We generally expect the
displacements of the Wigner crystal to be pinned either by the
periodic lattice potential or disorder (either is effective when
arbitrarily weak), so that this phase remains insulating.  As doping
increases from zero, the characteristic nodon momenta shift further
from $(\pm \pi/2, \pm \pi/2)$.  Continued doping to $\mu > \mu_{c2}$
leads to another transition into the ``normal'' state of the dual
theory, which is nothing but the d-wave SC.  Neglecting disorder, and
with weak lattice effects, quantum fluctuations are expected to drive
this $2+1$-dimensional flux lattice melting transition weakly first order.
For $\kappa \lesssim 1/\sqrt{2}$, one has instead {\sl type I} doping
at a single ``critical field'' $\mu_c$.  This is a first order
transition, accompanied by a jump in the hole concentration from $0$
to $x_c$ at $\mu=\mu_c$.  In the canonical ensemble with fixed
$0<x<x_c$, one expects two-phase coexistence, i.e a ``mixed'' phase in
the dual Ginzburg-Landau theory.  Taking into account long-range
Coulomb interactions, one arrives at the frustrated phase separation
physics discussed at length by Emery and Kivelson,\cite{EK}\ and all the
consequent issues.  Crude arguments (see Sec.~\ref{sec:NL}) suggest
$\kappa \sim 1-3$, in the weakly type II limit, but close enough to
the threshhold value to allow for different scenarios in different
materials.  Regardless, type II and some type I schemes imply NL {\sl
  phases} at finite doping before the onset of superconductivity.

An intermediate NL phase is extremely appealing from the theoretical
point of view, as it offers a compelling interpolation between the
undoped AF and the d-wave superconductor.  Consider the following
three important energies : the single-particle gap, $\Delta_1$, the
minimum energy required to add a charge $\pm e$ and spin $s=1/2$ to
the system; the two-particle gap, $\Delta_2$, the minimum energy to
add charge $\pm 2e$ and spin $s=0$; and the spin gap, $\Delta_s$, the
energy required to add spin $s=1$ but no charge.  In the AF, both
single- and two-particle gaps are non-zero, but the spin gap vanishes
due to low frequency magnons.  In the d-wave superconductor, the
two-particle gap vanishes, since the pairs have condensed, but spin
and single-particle gaps are ``almost'' non-zero (which we call
``$0^+$''), since only the quasiparticles carry these quantum numbers,
and their density of states vanishes with the energy.  Passing from
the SC to the NL, $\Delta_2$ changes from zero to non-zero, and
$\Delta_1$ changes from $0^+$ to a true gap.  The transition to the AF 
occurs then simply by developing a non-zero staggered magnetization.

This transition and other magnetic physics is discussed in
Sec.~\ref{sec:Neel}, using the N\'eel Lagrangian density
\begin{eqnarray}
  {\cal L}_N & = & {K_\mu \over 2}|\partial_\mu {\bf N}|^2  - V_N(|N|) 
\nonumber \\
  & & + g [{\bf N}\cdot \psi^\dagger
  \tau^z\sigma^y\bbox{\sigma} \psi^\dagger + {\rm h.c.}]  ,
  \label{eq:Neel}
\end{eqnarray}
where $g$ measures the strength of the coupling between the N\'eel order
parameter
and the nodons.
Eq.~\ref{eq:Neel}\ can be obtained by
introducing a $2k_{F}$ density-density interactions between the nodons
and decoupling the antipodal terms with the N\'eel vector ${\bf N}$.
Let us once again consider first the case of half-filling with
particle/hole symmetry.  For sufficiently strong interaction $g$, or
when the quadratic coefficient $r_N$ in $V_N(N)$ is negative, one
obtains an AF phase with $\langle{\bf N}\rangle \neq 0$.  In this
phase the nodons develop a gap and low-energy spin quanta are carried
entirely by spin waves.  Depending upon the ``mass'' $r_\Phi$ of $V_\Phi$,
this is either a simple AF or AF order coexisting with a d-wave SC.
Decreasing $g$ or increasing $r_N$ destroys the long-range AF
order and liberates the nodons.  This interesting phase transition is
discussed in Sec.~\ref{sec:Neel}.  Increasing $r_\Phi$ results in a further 
transition to the d-wave superconductor, which we believe is in the
three-dimensional inverted-XY universality class.  Tuning
$r_\Phi=r_N=0$ describes a multicritical point connecting directly
the AF and d-wave SC phases.

Without particle/hole symmetry, $\lambda \neq 0$, and another possible
phase exists: the coexisting AF and Nodal Liquid (AF/NL), with
long-range AF order at $(\pi,\pi)$ and power-law spin-density-wave
correlations from the nodons at incommensurate wavevectors.  This
phase may be difficult to distinguish experimentally from the pure AF,
and it seems possible that some of the well-studied {\sl undoped}
cuprate materials might well be in the AF/NL phase rather than the
pure AF.  In any event, the model of
Eqs.~\ref{eq:lagrangian},\ref{eq:Neel}\ provides a simple basis for
understanding the suppression of N\'eel order upon doping.  To see
this, assume that at half-filling $r_N > 0$.  Agreement with
experiment then requires that $g$ be sufficiently strong to induce AF
order.  As the chemical potential $\mu$ is increased above the charge
gap to induce holes into the system, the hole density or dual
``internal field'' becomes non-zero.  From Eq.~\ref{eq:lagrangian},
this creates an effectively larger particle/hole asymmetry
$\lambda_{\rm eff} - \lambda \sim x$.  This presents a competition.
By ordering the N\'eel vector the system can create a gap for the
nodons and reduce their kinetic energy.  However, at finite
$\lambda_{\rm eff}$ the nodon Dirac point would prefer to move away
from $(\pm\pi/2,\pm\pi/2)$, which reduces the gain in kinetic energy.
As $x$ increases, therefore, we may expect to drive transitions from
the AF to AF/NL and pure NL phases.  Of course, there are in fact many
different scenarios for type I and type II doping, small or large
intrinsic $\lambda$, etc.  These are discussed in Sec.~\ref{sec:Neel}.
Once magnetism has been discussed, we conclude with a summary of the
main points of the paper, open issues, relations to other work, and a
brief discussion of experimental implications.  Finally, two
appendices include technical details of microscopic and
$\epsilon$-expansion calculations.

\section{D-Wave and Duality}

\subsection{Model and Symmetries}

Consider a tight binding model of electrons hopping on a square 
lattice, with a local Hamiltonian satisfying certain general 
symmetries.  We will assume the system is both $U(1)$ and $SU(2)$ 
invariant (i.e. we neglect spin-orbit coupling), and has 
time-reversal, reflection, and four-fold rotational symmetry.  Sometimes
it will also be convenient to specialize to models
which possess an additional discrete particle/hole symmetry.
We denote lattice electron 
creation and annihilation operators as $c_{\alpha}^{\dagger}(\vec{x})$ 
and $c_{\alpha}^{\vphantom\dagger}(\vec{x})$, where $\vec{x}$ is the 
two-dimensional coordinate in the frame with $x=x_{1}$ and $y=x_{2}$ 
parallel to the $a$ and $b$ crystalline axes (i.e. to the $Cu$--$O$ bonds).
Here
$\alpha$ is a spin label.  In momentum space the kinetic energy
takes the usual form,
\begin{equation}
H_0 = \sum_{k \alpha} \epsilon_k c_{k\alpha}^\dagger c_{k \alpha}  ,
\end{equation}
and at this stage we allow for general electron interactions:
\begin{equation}
H_{int} = \sum_{k,q,k^\prime} V(k,q,k^\prime) c^\dagger_{-k+q,\alpha}
c^\dagger_{k\beta} 
c_{-k^\prime + q,\beta} c_{k^\prime,\alpha}   .
\end{equation}

A discrete particle/hole transformation is implemented by
\begin{equation}
c_\alpha (\vec{x}) \stackrel{p/h}{\longrightarrow}
e^{i\vec{\pi}\cdot\vec{x}}c_{\alpha}^{\dagger}(\vec{x}),
\end{equation}
with $\vec{\pi} = (\pi,\pi)$. Many common models (e.g. the Hubbard and 
t-J) are invariant under a particle/hole transformation at half-filling.   
Invariance of
the kinetic energy implies that $\epsilon_k = - \epsilon_{k+\vec{\pi}}$,
a form valid with near neighbor hopping.
However, a second
neighbor hopping term violates particle/hole symmetry.


As discussed in the introduction, we wish to describe the physics 
below the relatively strong d-wave pairing scale $T^*$, 
in order to approach the pseudo-gap phase from the superconducting 
side.  To do so, we imagine introducing a d-wave order parameter 
\begin{equation}
  \Delta_{k}(\vec{q})=\sum_{k^\prime}
V(k,q,k^\prime) \langle c_{-k^\prime+q\downarrow}
c_{k^\prime \uparrow} \rangle .
\end{equation}
BCS theory\cite{Schrieffer}\ can be implemented in terms of the spatially
varying pair field, obtained via Fourier transformation,
$\Delta_k (\vec{x})$.  The self-consistent gap equation is usually
solved for a {\it spatially uniform} order parameter, with
$\Delta_k \equiv \Delta_k(\vec{x})$.
Singlet pairing implies $\Delta_k = \Delta_{-k}$, and in
a d-wave superconductor $\Delta_k$ has four zero's or nodes
as $\vec{k}$ varies around the Fermi surface.
Our strategy will be to obtain an effective field theory which
has a {\it local} $d-$wave gap function, determined by strong coupling
physics below some length scale $\Lambda^{-1}$ of say 5-10 lattice spacings,
but which can fluctuate quantum mechanically on longer spatial scales.
These longer length scale quantum fluctuations will
be responsible for quantum disordering the d-wave
superconductor, and will allow us to access a new phase -- the nodal
liquid.  As we shall see, an important role is played by the
$d-$wave quasiparticles, which survive the quantum disordering.
To implement this approach, we first briefly recapitulate the properties of 
quasiparticles
in the $d-$wave superconductor.

\label{sec:duality}
\subsection{Quasiparticles}

With spatially uniform d-wave order given by $\Delta_{k}$,
the effective Hamiltonian for the 
quasiparticles is $H=H_{0}+H_{1}$, with $H_0$ the kinetic energy and
\begin{equation}
H_1 = \sum_k [ \Delta_k c_{k \uparrow}^\dagger c_{-k\downarrow}^\dagger
+ \Delta_k^* c_{-k\downarrow} c_{k\uparrow} ]    .
\end{equation}
Since $\Delta_k = \Delta_{-k}$ for singlet
pairing, it is natural to break sums into positive and negative $k_y$.
To do so, consider a four component fermion field,
$\Upsilon_{a\alpha}(\vec{k})$, at each wavevector with $k_y >0$ positive:
\begin{equation}
\Upsilon_{a\alpha}(\vec{k}) = \left[ \begin{array}{c}
\Upsilon_{11} \\ \Upsilon_{21} \\ \Upsilon_{12} \\ \Upsilon_{22} 
\end{array} \right] = \left[ \begin{array}{c}
c_{k\uparrow}^{\vphantom\dagger} \\ c_{-k\downarrow}^{\dagger} \\ 
c_{k\downarrow}^{\vphantom\dagger} \\ -c_{-k\uparrow}^{\dagger} 
\end{array}\right].
\end{equation}
In the second column vector the minus sign has been introduced
so that $\Upsilon_{a\alpha}$ transforms like a spinor under an $SU(2)$ 
rotation, i.e.
as $\Upsilon_{a\alpha} \rightarrow U_{\alpha \beta} \Upsilon_{a\beta}$.
Here $U = \exp(i \bbox{\theta} \cdot \bbox{\sigma})$ is a global
spin rotation with Pauli matrices $\bbox{\sigma}_{\alpha \beta}$.

In these variables, the quasiparticle Hamiltonian becomes 
\begin{equation}
H_{qp} = \left. \sum_k \right.^\prime \Upsilon^\dagger(\vec{k}) 
[ \tau^z \epsilon_k + \tau^+ \Delta_k
+ \tau^- \Delta^*_k ] \Upsilon(\vec{k})  ,
\end{equation}
where the prime on the summation denotes over $k_y$ positive, only,
and we have introduced a vector of Pauli matrices, $\vec{\tau}_{ab}$
acting in the particle/hole subspace.  Also, we
are employing the notation $\tau^{\pm} = (\tau^x \pm i \tau^y)/2$.

\begin{figure}
\hskip 1 cm
\psfig{figure=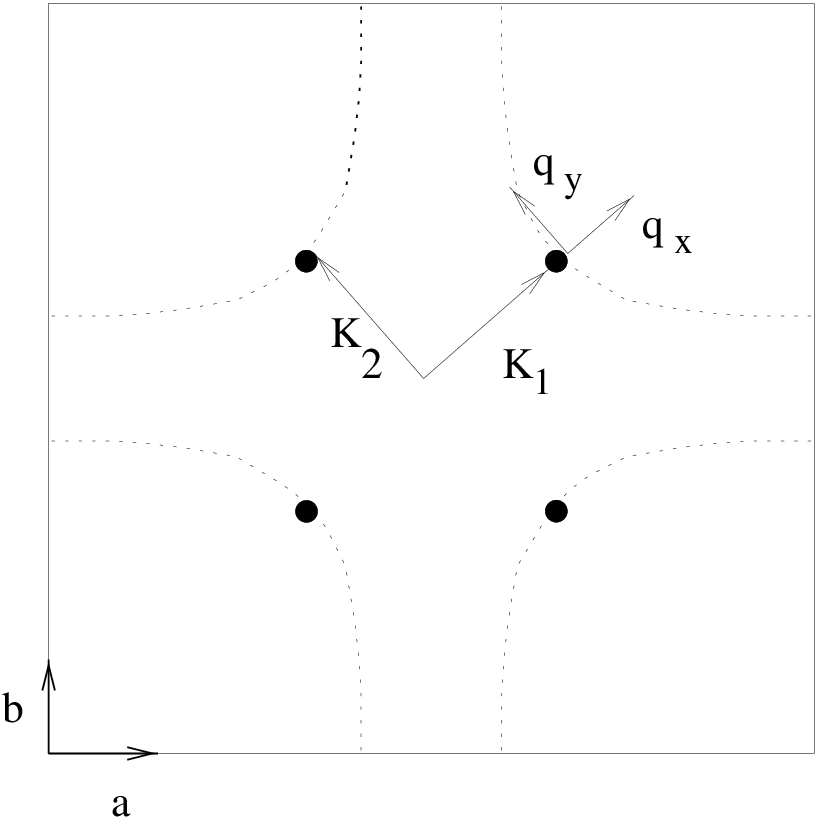,height=2.0in,angle=-90}
\vskip 0.5cm
{Fig.~2: The wavevectors $\vec{K}_i$, $q_x$, $q_y$
in relation to the $a$,$b$ axes. The dotted line
represents the putative Fermi surface.}
\end{figure}

With approximate particle/hole symmetry, the d-wave nodes are located 
near the special wavevectors $\pm \vec{K}_{j}$, with
$\vec{K}_{1} = (\pi/2,\pi/2)$ and 
$\vec{K}_{2} = (-\pi/2,\pi/2)$.  Since our aim is to obtain an effective
description at low energies and long lengthscales,
it is 
sufficient to focus on the gapless modes near these points, 
integrating out the electrons far away in the Brillouin zone.
It is then
convenient to introduce two continuum fields
$\Psi_j$, one for each pair of nodes, expanded around 
$\pm\vec{K}_{1},\pm\vec{K}_{2}$:
\begin{equation}
\Psi_{ja\alpha}(\vec{q}) = \Upsilon_{a\alpha}(\vec{K}_{j}+\vec{q}) .
\end{equation}
Here, the wavevectors $\vec{q}$ are assumed to be small, within
a circle of radius $\Lambda$ around the origin.
With this definition, the particle/hole transformation is extremely simple, 
\begin{equation}
\Psi  \rightarrow \Psi^{\dagger}.
\end{equation}
For this reason it is convenient to {\sl always} define the continuum
fields $\Psi$ around $\pm\vec{K}_{j}$, and account for deviations of
the node momenta from these values by a particle/hole
symmetry-breaking parameter $\lambda$.  

Once we have restricted attention to the momenta near the nodes, it is
legitimate to linearize in the quasiparticle Hamiltonian.  The
resulting continuum theory is more conveniently written in coordinates
perpendicular and parallel to the Fermi surface, so we perform the
rotation via $x \rightarrow (x-y)/\sqrt{2}$ and $y \rightarrow
(x+y)/\sqrt{2}$, correspondingly transforming the momenta $q_{x}$ and
$q_{y}$ (see Fig.~2).  Linearizing near the nodes, we put
$\epsilon_{K_1 + q} = v_F q_x$ where $v_F$ is the Fermi velocity and
$\Delta_{K_1 + q} = \tilde{\Delta} q_y$, where $\tilde{\Delta}$ has
dimensions of a velocity.  An identical linearization is possible
around the second pair of nodes, except with $q_x \leftrightarrow
q_y$.  Upon Fourier transforming back into real space,
$\Psi_j(\vec{q}) \rightarrow \Psi_j(\vec{x})$, we arrive at a compact
form for the Hamiltonian density of the quasiparticle excitations in
the d-wave superconductor: ${\cal H}_{qp} = {\cal H}_\Psi + {\cal
  H}_\lambda$ with
\begin{eqnarray}
\label{hpsi}
  {\cal H}_{\Psi} & = & \Psi_1^\dagger [v_F\tau^z i \partial_x +
  (\tilde{\Delta} \tau^+ + \tilde{\Delta}^* \tau^-) i \partial_y ] \Psi_1 
  \nonumber \\
  & & + (1 \leftrightarrow 2; x \leftrightarrow y)  ,
\end{eqnarray}
and the particle/hole symmetry breaking term,
\begin{equation}
{\cal H}_\lambda = \lambda \Psi_j^\dagger \tau^z \Psi_j  .
\end{equation}
The quasiparticle Hamiltonian takes the form of (four) Dirac equations,
and can be readily diagonalized giving
a dispersion relation
for the first pair of nodes, 
\begin{equation}
E_1(\vec{q}) = \pm \sqrt{ (v_F q_x + \lambda)^2 + |\tilde{\Delta}|^2 q_y^2} , 
\end{equation}
and a similar expression with $q_x$ and $q_y$ exchanged for the second
pair.  Notice that non-zero $\lambda$ indeed shifts the positions of
the nodes.


\subsection{Quantum Fluctuations}

Up to this point, we have taken a spatially constant gap function
$\tilde\Delta(\vec{x})$.  To disorder the d-wave superconductor it is
necessary to allow for quantum fluctuations of this order parameter.
It is tempting to uniformly suppress the complex order parameter, and
simply put $\tilde{\Delta} =0$.  But doing so recovers the
conventional metallic state with a Fermi surface.  Our task is
trickier, since we are searching for an {\sl intermediate} phase,
which has strong local $d-$wave pairing (which destroys the Fermi
surface) but with longer length scale quantum fluctuations destroying
the superconducting phase coherence.  Our task is similar to the
problem of describing the hexatic phase in a classical two-dimensional
triangular solid,\cite{Nelson79}\ which is intermediate between the
crystalline and liquid phases. Guided by this example and the
principle of pairing below $T^*$, we want to {\sl fix} the {\it
  magnitude} of the complex pair field, and introduce fluctuations of
its phase.

Pursuant to this
goal, we write
\begin{equation}
  \tilde{\Delta} \rightarrow v_{\scriptscriptstyle\Delta} e^{i\varphi}   ,
\end{equation}
where $v_{\scriptscriptstyle\Delta}$ is real and $\varphi$ can be
interpreted as the phase of the complex superconducting order
parameter.  The BCS gap equation has a degenerate manifold of
solutions, for arbitrary phase $\varphi$.  This degeneracy is
responsible for the Goldstone modes, wherein $\varphi$ varies slowly
in both space and time.  Our goal is to obtain an effective theory for
the space and time dependence of $\varphi$, similar in spirit to the
non-linear sigma models ``derived" for localization.\cite{Lee}\ 
Specifically, we focus on spatial variations of $\varphi(\vec{x})$ on
scales longer than $\Lambda^{-1}$.  Since $\varphi$ can vary
spatially, care is needed in introducing it into the quasiparticle
Hamiltonian:
\begin{equation}
  \tilde{\Delta} \tau^+ i\partial_y \rightarrow v_{\scriptscriptstyle\Delta} 
\tau^+ e^{i\varphi/2}
  (i \partial_y) e^{i\varphi/2}    .
\label{dgauge}
\end{equation}
This symmetric form leads to an hermitian Hamiltonian, physical
currents, and respects the symmetries of the problem. A careful
derivation of Eq.~\ref{dgauge}\ is given in
Appendix A. With
this prescription, the quasiparticle Hamiltonian becomes
\begin{eqnarray}
\label{hqp}
  {\cal H}_{qp} &= & \sum_{s=\pm} \Psi_1^\dagger [v_F\tau^z i \partial_x +
  v_{\scriptscriptstyle\Delta} \tau^s  e^{is\varphi/2}
  (i \partial_y) e^{is\varphi/2} ] 
  \Psi_1 \nonumber \\
  & & + (1 \leftrightarrow 2; x \leftrightarrow y)  .
\end{eqnarray}
Since $\varphi$ can also fluctuate with time, it will convenient
to consider the time dependence via a Lagrangian formulation.  The
Lagrangian density is
\begin{equation}
  {\cal L}_{qp} = \Psi_j^\dagger i \partial_t \Psi_j - H_{qp}   .
\end{equation}

The appropriate Lagrangian for the phase of the d-wave order parameter
is simply
\begin{equation}
\label{applag}
  {\cal L}_\varphi = {1 \over 2} \kappa_\mu  (\partial_\mu \varphi)^2  ,
\end{equation}
where the Greek index $\mu$ runs over time and two spatial
coordinates: $\mu = 0,1,2 = t,x,y$.  Here $\kappa_0$ is equal to the
compressibility of the condensate and $\kappa_j = - v_c^2 \kappa_0$
(for $j=1,2=x,y$) with $v_c$ the superfluid sound velocity.  We expect
that the pair compressibility $\kappa_0$ is approximately one half the
electron compressibility of the original electron model -- in the
absence of interactions.  If the pairing is electronic in origin, one
expects that the scale for the ``charge velocity" $v_c$ is the Fermi
velocity.

As discussed in the introduction, treatment of quantum phase
fluctuations is complicated by the mixing of particle and hole
variables via the complex gap function.  To isolate the uncertain charge, we
therefore perform a change of variables,
defining a new set of fermion fields $\psi_j$ via
\begin{equation}
  \psi_j = \exp(-i\varphi \tau^z/2) \Psi_j  .
  \label{transform}
\end{equation}
In the superconducting phase, and in the absence of quantum flucutations
of the order-paramater phase, one can set $\varphi = 0$,
and these new fermions are simply the d-wave quasiparticles.
However, when the field $\varphi$ is dynamical and fluctuates
strongly this change of variables is non-trivial.  In particular,
the new fermion fields $\psi$ are electrically {\sl neutral},
invariant under a global charge $U(1)$ transformation.  As we shall see,
when the d-wave superconductivity is quantum disordered,
these new fields will play a fundamental role, describing
low energy gapless excitations, centered at the former
nodes.  For this reason, we refer to these fermions as {\sl nodons}.
For completeness, we quote the symmetry properties of the nodon 
field under a particle/hole transformation.
Since $\varphi \rightarrow -\varphi$, one has simply
\begin{equation}
\psi \rightarrow \psi^{\dagger}.
\end{equation}

%

The full Lagrangian in the d-wave superconductor, ${\cal L} = {\cal
  L}_{\varphi} + {\cal L}_{qp}$, can be conveniently re-expressed in
terms of these nodon fields since ${\cal L}_{qp} = {\cal L}_{\psi} +
{\cal L}_{int} + {\cal L}_\lambda$ with a free nodon piece,
\begin{eqnarray}
  {\cal L}_{\psi} & = & \psi_1^\dagger [ i \partial_t - v_F \tau^z
  i\partial_x - v_{\scriptscriptstyle\Delta} 
  \tau^x i\partial_y ] \psi_1 \nonumber \\
  & & + (1 \leftrightarrow 2,x \leftrightarrow y)  ,
  \label{Lpsi}
\end{eqnarray}
interacting with the phase of the order-parameter:
\begin{equation}
  {\cal L}_{int} = \partial_\mu \varphi J_\mu .
\end{equation}
Here the electrical 3-current $J_\mu$ is given by
\begin{equation}
  J_0 = {1 \over 2} \psi_j^\dagger \tau^z \psi_j^{\vphantom\dagger}
  ,
  \label{eq:J0}
\end{equation}
\begin{equation}
  J_j = {v_F \over 2} \psi_j^\dagger \psi_j^{\vphantom\dagger}  .
  \label{eq:Jj}
\end{equation}
Because the transformation in Eq.~\ref{transform}\ is local, identical
expressions hold for these currents in terms of the quasiparticle
fields, $\Psi$.  The form of the particle/hole asymmetry term
remains the same in terms of the nodon fields:
\begin{equation}
{\cal L}_\lambda = \lambda \psi^\dagger_j \tau^z \psi_j .
\label{Llambda}
\end{equation}

It is instructive to re-express the components of
the currents $J_\mu$ back in terms of the original electron operators.  One finds
\begin{equation}
  J_0 = {1 \over 2} (c_{K_j}^\dagger c_{K_j}^{\vphantom\dagger} +
  c_{-K_j}^\dagger c_{-K_j}^{\vphantom\dagger}) , 
\end{equation}
(with an implicit spin summation) which corresponds physically
to the total electron density living at the nodes,
in units of the Cooper pair charge.  Similarly,
\begin{equation}
  J_j = {v_F \over 2} (c_{K_j}^\dagger c_{K_j}^{\vphantom\dagger} -
  c_{-K_j}^\dagger c_{-K_j}^{\vphantom\dagger}) 
\end{equation}
corresponds to the {\it current} carried by the electrons at the nodes.
Thus, $J_\mu$ can be correctly interpreted as
the quasiparticles three-current. 

To complete the description of a quantum mechanically fluctuating order
parameter phase interacting with the gapless fermionic excitations at the
nodes, we minimally couple to an external electromagnetic field,
$A_\mu$.  Since the nodon fermions are neutral, the only coupling is
to the order-parameter phase, via the substitution $\partial_\mu \varphi
\rightarrow \partial_\mu \varphi - A_\mu$.  For simplicity, here and
in the rest of the paper, we have set the Cooper pair charge $2e =1$.
The final Lagrangian then takes the form ${\cal L} = {\cal
  L}_{\varphi} + {\cal L}_{\psi} + {\cal L}_{int} + {\cal L}_\lambda$, with
\begin{equation}
  {\cal L}_{\varphi} = {1\over 2} \kappa_\mu (\partial_\mu \varphi - A_\mu)^2 ,
\end{equation}
\begin{equation}
  {\cal L}_{int} = (\partial_\mu \varphi - A_\mu ) J_\mu  ,
\end{equation}
and ${\cal L}_{\psi}$ still given by Eq.~\ref{Lpsi}.  

The time component of the electromagnetic field is proportional to the
chemical potential $\mu$, i.e. $A_0 = 2\mu$.  For electrons at
half-filling one has $\mu=0$.  Doping can be achieved by changing
$\mu$.  Long-ranged Coulomb interactions could be readily incorporated
at this stage by treating $A_0$ as a dynamical field and adding a term
to the Lagrangian of the form, ${\cal L}_{coul} = (1/2) (\partial_j
A_0)^2$.  The spatial components of the electromagnetic field, $A_j$,
have been included to keep track of the current operator.  In particular, the
total electrical 3-current is obtained by differentiating the
Lagrangian, i.e. $J^{tot}_\mu = \partial {\cal
  L} / \partial A_\mu$, which gives
\begin{equation}
  J^{tot}_\mu = \kappa_\mu
  (\partial_\mu \varphi - A_\mu) + J_\mu .
  \label{Jtot}
\end{equation}
Here the first terms are the Cooper pair 3-current, and the second
the quasiparticles current.  The equation of motion for the 
phase of the order-parameter, $\partial {\cal L}/\partial \varphi = 0$,
implies the continuity equation $\partial_\mu J^{tot}_\mu = 0$.

\subsection{Duality}

To quantum-disorder the d-wave superconductor, one must allow
for {\it vortices} in the pair-field phase, $\varphi$.  We do this
using field-theoretic duality, as described, e.g. in
Ref.~\onlinecite{fDuality}. 
To this end we introduce a vortex 3-current, $j^v_\mu$,
which satisfies,
\begin{equation}
  j^v_\mu = \epsilon_{\mu \nu \lambda} \partial_\nu \partial_\lambda
  \varphi.
  \label{jv}
\end{equation}
In the presence of vortices, $\varphi$ is multi-valued,
$\partial_\mu\varphi$ is not curl-free, and $j_\mu^v$ is non-vanishing.
In the desired dual representation, the vortices become the quantized
particles, rather than the Cooper pairs.  However, even in the dual
representation one still needs to conserve the total electrical
charge.  This can be achieved by expressing the total electrical
3-current as a curl,
\begin{equation}
  J^{tot}_\mu = \epsilon_{\mu \nu \lambda}  \partial_\nu a_\lambda  ,
  \label{gauge}
\end{equation}
where we have introduced a ``fictitious" dynamical gauge field,
$a_\mu$.  Upon combining Eqs.~\ref{Jtot}-\ref{gauge}, one can
eliminate the pair-field phase, $\varphi$, and relate
$a_\mu$ to the vortices:
\begin{equation}
  j^v_\mu  = \epsilon_{\mu \nu \lambda} \partial_\nu [ \kappa_\lambda^{-1}
  \epsilon_{\lambda \alpha \beta} 
  \partial_\alpha a_\beta +  A_\lambda - 
  \kappa_\lambda^{-1} J_\lambda ]  ,
  \label{eq:jva}
\end{equation}
where $J_\mu$ is the quasiparticle 3-current defined earlier in
Eqs.~\ref{eq:J0}-\ref{eq:Jj}.

A dual description is obtained by constructing a 
Lagrangian, ${\cal L}_D$, 
depending on $a_\mu$, $J_\mu$ and $j^v_\mu$, whose
equation of motion, obtained by differentiating with respect
to $a_\mu$, leads
to the above equation.
To assure that the vortex 3-current is conserved, it is useful
to introduce a 
complex field, $\Phi$,
which can be viewed as a vortex destruction operator.
Since a vortex acquires a $2\pi$ phase
upon encircling a Cooper pair, the vortex field should
be minimally coupled to $a_\mu$.
The appropriate dual Lagrangian can be conveniently decomposed as
${\cal L}_D = {\cal L}_\psi + {\cal L}_v + {\cal L}_a$,
where ${\cal L}_\psi$ is given in Eq.~\ref{Lpsi}.  
The vortex piece has the Ginzburg-Landau form,\cite{Tinkham}\
\begin{equation}
  {\cal L}_v = {\kappa_\mu \over 2} |(\partial_\mu - ia_\mu)\Phi|^2 - V_{\Phi}
  (|\Phi|)   ,
\end{equation}
where $\Phi$ is a (dimensionless) complex field for the vortices.
The vortex 3-current, following from 
$j^v_\mu = \partial {\cal L}_v /
\partial a_\mu$, is
\begin{equation}
  j^v_\mu = \kappa_\mu {\rm Im}[\Phi^*(\partial_\mu - ia_\mu)\Phi] .
\end{equation}
For small $|\Phi|$ (appropriate close to a second order transition)
one can expand the potential as, $V_\Phi(X) = r_\Phi X^2 + u_\Phi X^4$.
The remaining piece of the dual Lagrangian is
\begin{equation}
{\cal L}_a = {1 \over {2\kappa_0}} (e_j^2 - b^2) + a_\mu \epsilon_{\mu \nu 
\lambda} 
\partial_\nu (A_\lambda - \kappa_\lambda^{-1} J_\lambda)  ,
\label{Lgauge}
\end{equation}
with dual ``magnetic" and ``electric" fields: $b=\epsilon_{ij}
\partial_i a_j$ and $e_j = v_c^{-1} (\partial_j a_0 - \partial_0
a_j)$.  It is straightforward to verify that the dual Lagrangian has
the desired property that Eq.~\ref{eq:jva}\  follows from the equation of
motion $\partial {\cal L}_D / \partial a_\mu =0$.

\section{Nodal Liquid Phase}
\label{sec:NL}

In this section we employ the dual representation of
the d-wave superconductor to analyze the quantum disordered
phase - a new phase of matter which we refer to as a {\it nodal liquid}.
The dual representation comprises a complex vortex field,
which is minimally coupled to a gauge field, as well as a set
of neutral nodon fermions.  Without the nodons and in imaginary
time, the
dual Lagrangian is formally equivalent to
a classical three-dimensional superconductor at finite temperature,
coupled to a fluctuating electromagnetic field.
To disorder the d-wave superconductor, we must order
the dual ``superconductor" -- that is, condense the vortices.
The nature of the resulting phase
will depend sensitively on doping, since upon doping, the dual
``superconductor" starts seeing an applied ``magnetic field".
Below, we first consider the simpler case of half-filling.
We then turn to the doped case, where two scenarios are possible
depending on whether the dual ``superconductor" is Type I or Type
II.\cite{Tinkham}\

\subsection{Half-filling}

Specialize first to the case of electrons at half-filling, with particle-hole
symmetry.  In the dual representation,
the ``magnetic field", $b$, is equal to the
deviation of the total electron density 
from half-filling.
Thus at half-filling $\langle b \rangle = 0$ and the dual Ginzburg-Landau
theory is in zero applied field.
The quantum disordered phase corresponds to condensing the
vortices, setting $\langle \Phi \rangle = \Phi_0 \ne 0$.
In this dual Meissner phase, the vortex Lagrangian
becomes
\begin{equation}
{\cal L}_v = {1 \over 2} \kappa_\mu \Phi_o^2 a_\mu^2  .
\end{equation}
It is then possible to integrate out the field $a_\mu$.
The full Lagrangian in the nodal liquid phase
is then
\begin{equation}
{\cal L}_{nl} = {\cal L}_\psi + A_\mu I_\mu  +
{\epsilon_0 \over 2} E_j^2 - {B^2 \over {2 \mu_0}}+ O\left[(\partial 
J)^2\right]   ,
\label{Lnodliqu}
\end{equation}
where we have introduced the physical magnetic and electric fields:
$B = \epsilon_{ij} \partial_i A_j$ and $E_j = \partial_j A_0 - \partial_t A_j$.  
The last two terms describe a dielectric, with
magnetic permeability  $\mu_0 = \kappa_0 \Phi_0^2$
and dielectric constant $\epsilon_0 = (\mu_0 v_c^2)^{-1}$,
with the sound velocity entering, rather than the speed of light.
The external electromagnetic field is coupled to the 3-current
$I_\mu$, which can be expressed as a bi-linear of the nodon fermions as,
\begin{equation}
I_\mu = {\epsilon_0 \over {\kappa_0^2 v_c^2}} [ \kappa_\nu \partial_\nu^2
J_\mu - \kappa_\mu \partial_\mu (\partial_\nu J_\nu)] .
\end{equation}
Notice that this 3-current is automatically conserved:
$\partial_\mu I_\mu =0$.  

The order $(\partial J)^2$ terms which we have not written
out explicitly are quartic in the 
fermion fields, and also involve two derivatives.  Since ${\cal L}_\psi$
describes Dirac fermions in $2+1$ space-time dimensions,
these quartic fermion terms are highly irrelevant, and rapidly vanish
under a rescaling transformation.
Thus, in the absence of external electromagnetic fields,
the description of the nodal liquid phase is
exceedingly simple.  It consists of four neutral
Dirac fermion fields --  two spin polarizations ($\alpha = 1,2$) for each of 
the two pairs of nodes.

Despite the free fermion description, the nodal liquid phase is highly
{\it non-trivial} when re-expressed in terms of the underlying
electron operators.  Indeed, the $\psi$ fermion operators are built
from the quasiparticle operators $\Psi$ in the d-wave superconductor,
but are electrically neutral, due to the ``gauge transformation" in
Eq.~\ref{transform}.  Moreover, in the nodal liquid phase, the Cooper pairs are
{\it not} superconducting, but rather in a dielectric Mott-insulating
phase,\cite{Fisher89}\ immobilized by their commensurability with the
underlying crystal 
lattice.  Although the $\psi$ fermions are electrically neutral, they
do carry a new conserved ``charge".  In fact, there are {\it four} new
conserved charges, since the Lagrangian is invariant under the global
transformations $\psi_{ja\alpha} \rightarrow e^{i\theta_{j\alpha}}
\psi_{ja\alpha}$ for arbitrary constant phases, $\theta_{j\alpha}$,
with $j=1,2$ and $\alpha = 1,2$.  We refer to the $\psi$ fields as {\it 
nodon} operators, their quanta as {\sl nodons}, and the associated 
conserved quantities as ``nodon charges".  As seen from Eq.~\ref{Lpsi}, 
these conserved charges
are related to the quasiparticle current, since $J_j = (v_F/2)
\psi_j^\dagger \psi_j$.  However, in the nodal liquid phase the
electrical current operator is $I_\mu$, not $J_\mu$, since $I_\mu =
\partial {\cal L}_{nl}/ \partial A_\mu$.

\subsubsection{Spin response}

The spin response functions in the nodal liquid are rather straightforward,
since the electron spin operators have a simple representation
in terms of the nodons.  In particular, consider
the spin operator for small momentum,
\begin{equation}
{\bf S}_q = {1 \over 2} \sum_k c^\dagger_{k + q} \bbox{\sigma} 
c_{k}^{\vphantom\dagger} .
\end{equation}
At low energies in the nodal liquid phase one can focus on momenta
near the nodes:  $\vec{k}=\pm \vec{K}_j$.
The electron operators near the nodes
can be rewritten in terms of the nodon operators,
and one finds that back in real
space the long-wavelength piece of the spin operator, 
${\bf S}(\vec{x}) = \sum_q \exp(i \vec{q} \cdot 
\vec{x}) {\bf S}_q$, is simply
\begin{equation}
{\bf S} (\vec{x}) = {1 \over 2} \psi^\dagger_{ja}(\vec{x}) \bbox{\sigma} 
\psi_{ja}^{\vphantom\dagger} (\vec{x})  .
\end{equation}
Spin correlation and response functions can then be computed
from the free nodon theory.  For example, the uniform 
spin susceptibility is given by
\begin{equation}
\chi = \int_0^\infty \!\! dE (-\partial f/\partial E) \rho_n(E)  ,
\end{equation}
where the nodon density of states is $\rho_n(E) = (const) E/v_F 
v_{\scriptscriptstyle\Delta}$,
and $f(E)$ is a Fermi function.  One finds $\chi \sim T/v_F 
v_{\scriptscriptstyle\Delta}$.
There are also low energy spin excitations at wavevectors which
span between two different nodes.  The associated spin operators
can readily be be re-expressed in terms of the nodon fields.
For example, the staggered magnetization operator, ${\bf 
S}_{\vec{\pi}}$, is
\begin{equation}
  {\bf S}_{\vec{\pi}} = {1 \over 2} \left[\psi^\dagger (\tau^y
  \bbox{\sigma} \sigma^y)  
  \psi^\dagger + {\rm h.c.}\right] .
  \label{Spi}
\end{equation}
Notice that this operator is actually ``anomalous" in terms of the conserved
nodon charge.  We will return to the effects of 
finite wavevector magnetic fluctuations and ordering in Section IV.

In addition to carrying spin, the nodons carry energy, and so
will contribute to the thermal transport.  
At finite temperature, Umklapp scattering processes (or impurities)
give a finite thermal conductivity;
in their absence the nodon thermal conductivity
is infinite.

\subsubsection{Charge response}

The electrical charge properties in the nodal liquid
phase are, however, somewhat trickier.
Imagine changing the chemical potential
away from $\mu=0$.  In terms
of the dual vortex ``superconductor" this corresponds to
applying an external ``magnetic" field,
due to the coupling ${\cal L}_\mu = -2\mu b$.
The vortices, however, are in the ``Meissner" phase,
and for $\mu \le \mu_c$ the applied field will be screened
out, maintaining the internal field at $b=0$.  
That is, the electron density will
be pinned at half-filling, until the chemical potential
passes through the Mott gap for the insulating system
of Bosons (Cooper pairs).\cite{Fisher89}\  

Despite the presence of a charge gap, there are low energy
current fluctuations in the nodal liquid.  Indeed,
in this phase the electrical current operator
is $I_\mu$, which is bi-linear in terms
of the nodon fermions, $\psi$.
One might
imagine employing this current operator to compute the electrical
conductivity in the nodal liquid.  For this, one requires
computing a two-point correlator of the
electrical current operator at zero wavevector (say in the $x-$direction)
$I_x(q=0) = (\epsilon_0/\kappa_o v_s^2) \partial_t^2 J_x(q=0)$.
But notice that $J_x(q=0)$ is proportional to a globally conserved
nodon charge, since
$J_x(\vec{x})  = (v_F/2) \psi_1^\dagger \psi_1^{\vphantom\dagger}$.
Thus, when the nodon number is conserved one has $I_x(q=0)=0$,
and the nodons {\it do not} contribute
to the electrical conductivity.  (There will of course
be a response at finite frequencies in the imaginary
part of the conductivity from the
Mott-insulating phase of the Cooper pairs.)

When impurities or Umklapp scattering is present, however,
the nodon number is no longer conserved, and
the nodons presumably will contribute to the real
part of the electrical conductivity,
at least at finite frequencies.
Specifically, the umklapp scattering term with momentum transfer
$2\vec{\pi}$ is given by
\begin{equation}
{{\cal L}_{umklapp}} = u\, {\epsilon^{ABCD}}
{\psi_{jA}}{\psi_{jB}}{\psi_{jC}}{\psi_{jD}} + {\rm h.c.},
\end{equation}
where the composite index $A$ runs over $1,2,3,4$
corresponding to $a\alpha = 11,12,21,22$. By
power counting, this term is irrelevant by
one power of frequency. Hence, it enters
scaling forms in the combination
${u_{eff}}=u\omega$.

According to the Kubo formula,
\begin{eqnarray}
\sigma(\omega) &\sim& \frac{1}{\omega}\,
\langle{I_x(q=0,\omega)} {I_x(q=0,-\omega)}\rangle\cr
&\sim& {\omega^3}\, \langle{J_x(q=0,\omega}) {J_x(q=0,-\omega)}\rangle.
\end{eqnarray} 
From scaling, we expect the latter correlation
function to vary as $\omega$. However, as
we noted above, it actually vanishes in the
absence of Umklapp scattering; therefore
it is determined by the correction to scaling,
which, naively, is of the form:
\begin{equation}
\sigma(\omega) \sim {\omega^3}\cdot\omega\cdot{u_{eff}^2}
\sim u^2{\omega^6}.
\end{equation}

Finally,
it is instructive to consider the behavior of the
electron Green's function, which can be accessed
in photo-emission and tunneling experiments.  The electron operator
$c_\alpha(\vec{x})$ can be conveniently
decomposed in terms of the nodon operators by focusing
on momenta near the nodes.
For example, near the node at $\vec{K}_j$
one can write,
\begin{equation}
c_\alpha(\vec{x}) = e^{i \vec{K}_j \cdot \vec{x}} 
e^{i\varphi(\vec{x})/2}
\psi_{j1\alpha} (\vec{x})  + ...
\end{equation}
In the nodal liquid phase,
the electron Green's function, $G(\vec{x}, t) = \langle c^\dagger(\vec{x},
t) c(\vec{0},0) \rangle$ factorizes as,
\begin{equation}
G(\vec{x}, t) = e^{i \vec{K}_j \cdot \vec{x}} \langle 
e^{-i\varphi(\vec{x},t)/2} e^{i \varphi(\vec{0},0)/2} \rangle
\langle \psi^\dagger_{j1\alpha}(\vec{x},t) \psi_{j1\alpha}(\vec{0},0) \rangle
.
\end{equation}
Although the nodon correlator is a power law, falling off as
$|x|^{-2}$ and $t^{-2}$, one expects the correlator over exponentials
of the pair field phase to fall off exponentially in the nodal liquid,
since the Cooper pairs (Bosons) are locked in a Mott insulating phase.
This indicates a gap in the electron spectral function at the Fermi
energy, of order $\Delta_1 \sim \mu_c$ at the nodes.  If $\mu_{c}$ is
small relative to $T^{*}$, the corresponding gap will show strong
four-fold anisotropy in momentum space, varying from of order $T^{*}$
down to of order $\mu_{c}$ near the nodon wave-vectors.  In the
discussion section, we comment briefly on how such a d-wave pseudo-gap
feature is likely enhanced when the NL is doped in the presence of
impurities.

\subsection{Doping the Nodal Liquid}

We now consider the effects of doping charge into
the nodal liquid phase.  In a grand canonical ensemble
this is achieved
by changing the chemical potential, $\mu = A_0/2$.  
In the dual Ginzburg-Landau description of the vortices,
a chemical potential acts as an applied dual field, as seen
from Eq.~\ref{Lgauge}, since 
\begin{equation}
{\cal L}_\mu = - 2 \mu b   .
\end{equation}
The dual magnetic field, $b = \epsilon_{ij} \partial_i a_j$,
is the total electric charge in units of $2e$.
For a hole doping with concentration $x$, one has $\langle b \rangle =
x/2a_0^2$,  
with $a_0$ the crystal lattice constant.
Provided the applied dual field, $2\mu$, is smaller
than the critical field ($2\mu_c$) of the Ginzburg-Landau theory,
the dual superconductor stays in the Meissner phase -- which
is the nodal liquid phase at half-filling.  But for
$\mu \ge \mu_c$ dual flux will penetrate the Ginzburg-Landau superconductor,
which corresponds to doping the nodal liquid phase.  The form
of the dual flux penetration will depend critically on
whether the dual Ginzburg-Landau theory is Type I or Type II.
Within a mean-field treatment this is determined
by the ratio of the dual penetration length, $\lambda_v$,
to the dual coherence length, $\xi_v$ (where the subscript
$v$ denotes vortices).  In particular, Type II behavior
is expected if $\lambda_v/\xi_v \ge 1/\sqrt{2}$, and Type I
behavior otherwise.  In the Ginzburg-Landau description
$\lambda_v$ determines the size of a dual flux tube,
which is simply a Cooper pair.  We thus expect
that $\lambda_v$ will be roughly equal to
the superconducting coherence length, $\xi$, which is 
perhaps $15-20 \AA$ in the cuprates.  On the other
hand, $\xi_v$ is the size of the ``vortex-core" in the dual
vortex field, and  presumably can be no smaller than the microscopic
crystal lattice spacing, $\xi_v \ge 3-5\AA $.
This reasoning suggest that $\lambda_v/\xi_v$ is probably
close to unity, so that either Type I or Type II behavior
might be possible - and could be material dependent.
We first consider
such Type II doping, returning below to the case of
a Type I Ginzburg-Landau theory.

\subsubsection{Type II Behavior}

The phase diagram of a clean three-dimensional type II superconductor
is well understood.\cite{Tinkham}\ Above the lower critical field,
$H_{c1}$, flux tubes penetrate, and form an Abrikosov flux lattice -
usually triangular.  As the applied field increases the flux tubes
start overlapping, when their separation is closer than the
penetration length.  Upon approaching the upper critical field
$H_{c2}$ their cores start overlapping, the Abrikosov flux lattice
disappears, and the superconductivity is destroyed.  Mean field theory
predicts a second order transition at $H_{c2}$, but with thermal
fluctuations one expects this to become weakly first
order.\cite{Nelson88}\ This weak first order transition separates a
flux-lattice phase from a non-superconducting flux-liquid.

These results hold equally well
for our dual Ginzburg-Landau superconductor, except
that now the direction parallel
to the applied field is actually imaginary time.
Moreover, the
Ginzburg-Landau order parameter
describes quantum vortices, and the penetrating flux
tubes are Cooper pairs. 
Upon doping the nodal liquid with $\mu > \mu_{c1}$, charge is added
to the 2d system,
which corresponds to the penetration of dual magnetic flux. 
In this dual transcription, the resulting Abrikosov flux-lattice phase
is a Wigner crystal of Cooper pairs, with one Cooper pair
per real space unit cell of the lattice.  
We denote this paired Wigner crystal phase by PWC.
Upon further doping, one passes via a weak first
order transition (at $\mu = \mu_{c2}$) into the dual flux-liquid phase.
In this phase the lattice of Cooper pairs has melted, and they are
free to condense - this is the d-wave superconductor.
This latter transition should occur when the spacing between
dual flux tubes becomes roughly comparable to the coherence length,
$\xi_v$.  Experimentally, superconductivity typically sets in
for $x = 0.1$, which corresponds to one
Cooper pair for every 20 or so Cu atoms, and a mean pair separation
of 4-5$a_0$.  This again suggests that
$\lambda_v/\xi_v$ is probably of order one.

In the Cooper pair Wigner crystal phase, translational symmetry is
spontaneously broken.  However, in a real material the Wigner crystal
will have a preferred location, determined by impurities and perhaps
crystal fields, which will tend to pin and immobilize the Wigner
crystal.  The resulting phase should be an electrical insulator.
Moreover, in two-dimensions even weak impurities will smooth the
weakly first order transition between the Wigner crystal and
superconducting phases.  (In the absence of impurities, long-ranged
Coulomb interactions preclude phase separation, so a mixed phase would
result - see Type I behavior below.)

A striking and unusual feature of the PWC phase,
is that it {\it co-exists} with the nodal liquid, as we now argue.
With a weak (commensurate) pinning potential present,
the Wigner crystal phase is
a dielectric.  The charge response is
thus essentially the same as that of the 
undoped phase at half-filling, except with a modified
dielectric constant and magnetic permeability.
Thus even with doping, it is possible to
``integrate out" the charge fluctuations described
by the fields $a_\mu$, and arrive at the nodal liquid
Lagrangian Eq.~\ref{Lnodliqu}, except with different values
of $\epsilon_0$ and $\mu_0$.
The only complication is that there will be a background
frozen in charge density, from the Wigner crystal of Cooper pairs, so that 
$\langle b(\vec{x}) \rangle = n(\vec{x}) = n_0 + \delta n(\vec{x})$.
Here, the mean pair density is simply $n_0 = x/2a_0^2$ for doping $x$,
and $\delta n(\vec{x})$ has the periodicity of the Wigner crystal.
This ``background" field couples to the nodons,
and from Eq.~\ref{Lgauge}\ leads to a term of the form,
\begin{equation}
{\cal L}_b = \langle b(\vec{x}) \rangle \kappa_0^{-1} J_0 = {1 \over {2 \kappa_0}} 
n(\vec{x})
\psi^\dagger_j \tau^z \psi_j  .
\end{equation}
What is the effect of this term on the nodons?  Consider
first the spatially constant piece, proportional to $n_0$.
This term can be absorbed into ${\cal L}_{\psi}$,
which contains a term
of the form, $\psi_j^\dagger v_F q_j \tau^z \psi_j$,
and leads to a momentum space shift of the nodes,
with $q_j \rightarrow q_j + (n_0/2v_F \kappa_0)$.  Since
the pair compressibility $\kappa_0 \sim \partial n_0/\partial \mu$,
the shift satisfies $v_F \delta q \sim \delta \mu$,
as expected from the change of the area enclosed
by the Fermi surface upon doping.

The spatially varying background density, $\delta n(\vec{x})$, which
has the periodicity of the underlying Wigner crystal, causes a mixing
between nodon states at momentum differing by Wigner crystal
reciprocal lattice vectors.  When the reciprocal lattice vectors are
{\it larger} than the momentum cutoff, $\Lambda$, of the nodons, the
$\delta n(\vec{x})$ term cannot scatter within the low energy nodon
theory, and can be dropped.  At lower pair densities, it becomes
necessary over a range of length scales to retain the new periodicity
by working with nodon Bloch states, rather than plane waves.  (Since
only linear derivatives enter into the nodon Lagrangian, Bloch
wavefunctions in the lowest band can be readily constructed.)  In any
event, at length scales larger than the Wigner crystal lattice
spacing, the form of the effective theory of the nodons is identical
to that at half-filling.

We thereby arrive at a description of a rather remarkable new phase of matter.
A Paired Wigner Crystal (PWC) of doped Cooper pairs co-exists with neutral
gapless fermionic excitations -- the nodons.  In this co-existing phase,
which we denote as PWC/NL, low energy spin and thermal
properties will be dominated by the nodons.  The behavior will
be qualitatively similar to that in the undoped nodal liquid phase.
We propose that this PWC/NL phase is present in the pseudo-gap
region of the high $T_c$ cuprates.

\subsubsection{Type I Behavior}

In a classical Type I superconductor, the applied field is expelled
until the critical $H_c$ is exceeded.\cite{Tinkham}\  At this point there
is a {\it first order} phase transition
from the Meissner phase with all the flux expelled,
to a normal metal phase in which (essentially) all
the field penetrates.  If a thin film type I superconductor
is placed in a perpendicular field, screening currents are unable
to expel all the flux, and a ``mixed" or
``intermediate" state occurs.   In this mixed phase, regions of
superconductivity co-exist with normal metallic regions.
In some cases the superconducting regions form stripes,
but generally the lowest energy configurations are determined
in large part by material imperfections, and tend to be ``history"
dependent.

If our dual Ginzburg-Landau theory describing quantized vortices
is of type I, then similar properties are expected.
Specifically, as the chemical potential increases,
the dual field -- which is the Cooper pair density -- remains
at zero until a critical chemical potential $\mu_c$ is
reached.
At this point there is a first order phase transition,
between the nodal liquid phase at half-filling, and
a d-wave superconductor at finite doping, $x_c$.
At fixed doping $x < x_c$, phase separation is
impeded by long-ranged Coulomb interactions between the Cooper pairs.
The system will break apart into co-existing ``micro-phases"
of nodal liquid and d-wave superconductivity.
The configuration of the
co-existing ``micro-phases" will
be determined by a complicated competition between 
the Coulomb energy and the (positive) energy of the
domain walls.  In practice, impurities will also probably
play a very important role.  This doping scenario is
similar to that envisaged by Emery and Kivelson,\cite{EK}\ who have
extensively discussed the possibility of
phase separation as a mechanism for high $T_c$ superconductivity.
Unfortunately, with the transition being strongly first order
in this case, the associated physics is rather non-universal.

\section{Antiferromagnetism in the Nodal Liquid}
\label{sec:Neel}

\subsection{Effective Action for Antiferromagnetism}

We now turn to the low-doping region of the phase diagram of
Fig.~1. Retaining the Nodal Liquid as the underlying
description of the low-energy fermionic degrees of freedom, we
consider antiferromagnetic ordering.  In principle, this can arise in
two ways.  Antiferromagnetism could stem from interactions between
nodons, i.e. physics below the scale $T^*$.  This can be modeled in
principle by including simple inter-nodon interactions.  The
experimental coincidence of $T^*$ with the N\'eel temperature and
magnon bandwidth $J$ suggests that such a separation of scales is not
valid. Instead, antiferromagnetic correlations may exist already at
(high) energies comparable to $T^*$.  This sort of local AF amplitude
could be captured by decoupling spin-spin (or, e.g. on-site Hubbard)
interactions in a microscopic model with a Hubbard-Stratonovich
transformation.  Such a decoupling introduces a conjugate field ${\bf
  M}$, which interacts with the electron operators via a term of the
form $H_{M} = \sum_{\vec{x}} {\bf M}(\vec{x}) \cdot c_\alpha^\dagger {1
  \over 2} \bbox{\sigma}_{\alpha\beta} c_\beta^{\vphantom\dagger}$.
Integrating out high-energy degrees of freedom generates an effective
action for ${\bf M}$.  We expect dominant ordering tendencies at
momentum $\vec{\pi} = (\pi,\pi)$, and so decompose
\begin{equation}
  {\bf M} \sim {\bf M}_0 + 
\exp(i\vec{\pi}\cdot\vec{x}) {\bf N},
\end{equation}
where ${\bf N}$ and ${\bf M}_0$ are slowly-varying.  The fields ${\bf
  M}_0$ and ${\bf N}$ have the physical interpretation of
the coarse-grained uniform and staggered magnetization.  Focusing on
the N\'eel ordering, we imagine integrating out ${\bf M}_0$ to obtain
the Lagrangian
\begin{eqnarray}
\label{Lneel}
{\cal L} = {1 \over 2}K_{\mu} |\partial_{\mu}{\bf N}|^{2} - 
V_{N}(|N|) + g {\bf N}\cdot {\bf S}_{\vec{\pi}},
\end{eqnarray}
where $K_{0} = K$, $K_1 = K_{2} = -v_{s}^{2} K$, with $v_{s}$ the 
spin-wave velocity in the AF.  Here ${\bf S}_{\vec{\pi}}$ 
is the spin operator at momentum $\vec{\pi}$, expressed as 
a bi-linear in terms of the electron operators as in Section III.
The staggered magnetization operator
can be readily re-expressed in terms of the nodons as,
\begin{equation}
{\bf S}_{\vec{\pi}} = {1 \over 
2}\left[\psi^{\dagger}\tau^{y}\bbox{\sigma}\sigma^{y}\psi^{\dagger} + 
{\rm h.c.}\right],
\end{equation}
given earlier in 
Eq.~\ref{Spi}.  Near any phase transitions, and for most 
phenomenological purposes, it is sufficient to take a simple form for 
the potential: $V_N(|N|) = r_{N}|N|^{2} + u_{N}|N|^{4}$.  The 
parameter $r_{N}$ controls the presence or absence of AF order.  In 
mean-field theory, and neglecting for the moment the nodon coupling 
$g$, the ground state passes from long-range to short-range AF order 
as $r_{N}$ is tuned from negative to positive.  
A precise determination of $r_N$, $u_N$, and $g$
in terms of $t$, $t'$, $U$, $x$, etc. is the province of a
more microscopic theory.

We note that, in principle, Eq.~\ref{Lneel}\ allows for the 
possibility of {\sl incommensurate} spin-density-wave ordering at wavevectors 
other than $(\pi,\pi)$, which would correspond to a state which 
spontaneously develops a spatially periodic expectation value for 
${\bf N}$.  We find such ordering unlikely within this model, 
however, and have therefore {\sl not} included terms responsible for 
locking in possible higher-order commensurate magnetic wavevectors.  
Since incommensurate order seems not to be realized experimentally at 
low doping, we hope this omission is unimportant.

\subsection{Magnetism and Phases at Half-Filling}

Once we have coupled in the N\'eel order parameter field,
we can describe magnetic phases, in addition to
the nodal liquid and d-wave superconducting phases of earlier sections.
Here we first focus on the situation at half-filling, where
our effective field theory already 
descibes a number of magnetic and non-magnetic ground states.
It will be useful to further specialize initially to
models with particle-hole symmetry,
returning later to the half-filled but particle-hole
asymmetric case below.

\subsubsection{Particle-hole symmetric case}

The full effective Lagrangian has two order parameter fields,
the N\'eel order parameter, ${\bf N}$, and the vortex complex field,
$\Phi$, which is minimally coupled to a gauge field, $a_\mu$.
The N\'eel order parameter is directly coupled to the nodons,
whereas the vortex field only sees the nodons indirectly
via the gauge field.  Ordering of the two fields
is determined by the coefficients of 
the quadratic terms in the Lagrangian, namely $r_N$ and $r_\Phi$.
It will be convenient to plot the phase diagram at
half-filling in the $r_N - r_\Phi$ plane.  The phase diagram
with particle-hole symmetry ($\lambda =0$) is shown in Fig.~3a.
Here we briefly discuss each of the four phases.

\begin{figure}
\psfig{figure=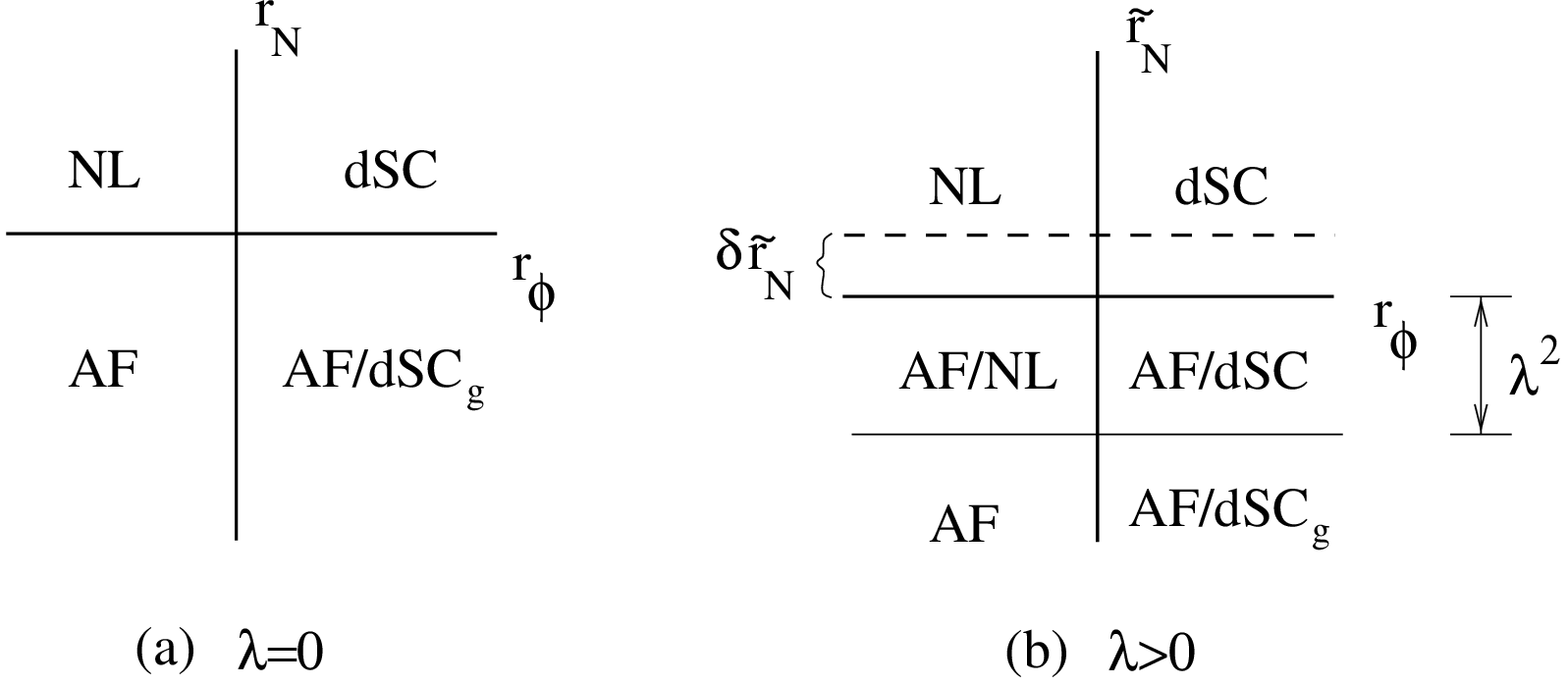,height=1.5in,angle=-90}
\vskip 0.5cm
{Fig.~3: Phase diagrams at half-filling for the
particle-hole (a) symmetric and (b) asymmetric cases.  In (b), both
horizontal phase boundaries shift downward with increasing
particle/hole asymmetry parameter $\lambda$, diminishing the domain of 
the AF phase.}
\end{figure}

Consider first $r_N$ large and positive, so that
N\'eel order is not present.  For $r_\Phi$ negative, the vortices
will condense (the Meissner phase in the dual Ginzburg-Landau theory)
leaving the nodons as the only low energy gapless excitations.
This is the nodal liquid phase.  As $r_\Phi$ changes sign,
the vortices will disorder, entering the ``normal" phase
of the dual Ginzburg-Landau theory.  This phase corresponds
to the d-wave superconductor, shown in the upper right quadrant
of Fig.~3a.  The d-wave superconducting phase
can be obtained in a microscopic lattice model even at
half-filling, by appropriately choosing 
the electron interaction
terms.\cite{Assaad}\

When $r_N$ is large and negative, the model magnetically orders
into the antiferromagnetic N\'eel phase.  With particle-hole symmetry
the antipodal nodes are separated by the N\'eel ordering wavevector,
$(\pi,\pi)$, so the nodons are ``nested".  This 
opens a gap in the nodon spectrum, as we now demonstrate.
With either sign of $r_\Phi$ the gauge field $a_\mu$ can
be integrated out, generating irrelevant four fermion nodon interaction
terms, which can be dropped. 
Then, upon putting $\langle {\bf N} \rangle = N_0 \hat{\bf{y}}$ into
the effective Lagrangian, we arrive
at the following quadratic Lagrangian for the nodon fields,
\begin{equation}
  {\cal L}_{nodon} = {\cal L}_\psi + gN_0 (\psi^\dagger_j \tau^y
  \psi^\dagger_j +  {\rm h.c.})    .\label{Lnodon}
\end{equation}
This model can be readliy diagonalized with an appropriate
Bogoliubov transformation, giving energy eigenvalues,
\begin{equation}
E_1(\vec{q}) = \pm \sqrt{ (v_F q_x)^2 + (v_{\scriptscriptstyle\Delta} q_y)^2 + 
(gN_0)^2 }  ,
\end{equation}
in the $j=1$ sector, and an identical form with $q_x$ and $q_y$
interchanged for the other pair of nodes ($j=2$).  In all
nodon sectors there is a non-zero {\it gap},
equal to $g N_0$.  In the lower left quadrant of Fig.~3a, with
$r_\Phi$ negative, this corresponds to the
usual N\'eel antiferromagnet.  With the nodons
gapped out, the only low energy excitations are the
spin waves of the antiferromagnet.  For $r_\Phi$ positive,
in the lower right quadrant, antiferromagnetism co-exists
with d-wave superconductivity.  In the
d-wave superconductor, the nodons become equivalent
to the d-wave quasiparticles, so that the
d-wave state in this quadrant is rather unusual.
In particular, it is a d-wave superconductor
with a full single particle gap,
and an absence of nodal quasiparticles.
We have denoted this with a subscript $g$ -- for gap -- in the
phase of the lower right quadrant.

Before turning to the effects of particle-hole asymmetries,
it is interesting to briefly discuss the nature of the
phase transitions between the four phases in Fig.~3a.
Consider first the vertical phase boundary, separating
the superconducting from non-superconducting phases.
For $r_N$ negative there are no gapless nodons, and the magnet
is ordered in both phases.  At the transition,
$r_\Phi =0$, where superconductivity develops, we can 
employ the dual Ginzburg-Landau theory.
Equivalently, 
we can return to the original representation (before duality)
in terms of the pair field phase, $\varphi$,
which (in the absence of long-ranged Coulomb
interactions\cite{Fisher88}) is simply
the classical three-dimensional XY model.  The resulting
transition is in the classical $3d$-XY universality class.
For $r_N$ positive, we have to worry
about the presence of gapless nodons, which might effect
the nature of the superconducting transition at $r_\Phi =0$.
However, since the dual Ginzburg-Landau field $\Phi$ is
only indirectly coupled to the nodons, via the gauge field
$a_\mu$, we expect that the transition
will still be in the $3d$-XY universality class.  Indeed,
power counting in $d$ spatial dimensions
about the Gaussian fixed point, reveals
that the coupling term (of the form $\partial a \psi^\dagger \psi$)
has scaling dimension $(3d+1)/2$.  Being greater than $d+1$
for all $d \ge 1$, and much greater near the upper-critical
dimension ($d_{uc} = 3$), one expects this coupling to 
be strongly irrelevant in the two-dimensional case of interest.

Consider next the phase transition into the antiferromagnetic state,
upon crossing the horizontal axis.  For either sign of
$r_\Phi$, since the vortices and gauge field $a_\mu$ can
be integrated out generating irrelevant four-nodon interactions,
the transition is described by the N\'eel ordering field
coupled to the nodon bi-linear, with Lagrangian
${\cal L} = {\cal L}_\psi + {\cal L}_N$.
This is an interesting field theory describing Dirac fermions
anomalously coupled to a fluctuating $O(3)$ field.  Power counting
about the Gaussian fixed point (free nodons and $r_N = u_N =0$)
reveals that the coupling term is relevant in two dimensions, but 
if the model is suitably generalized
to higher dimensions becomes marginal in $d=3$.  This suggests
an attack 
near four space-time dimensions, working
perturbatively for small $\epsilon = 4 - (d +1)$.  A complication
is that there are in fact three independent velocities,
two for the nodons ($v_F, v_{\scriptscriptstyle\Delta}$) and one for the
N\'eel spin field ($v_s$).  In Appendix B we first consider the
special case in which all three velocities are set equal.
This model is then a bona fide relativistic field theory,
with Lorentz invariance.  A leading order perturbative
renormalization group in $\epsilon$ reveals the existence
of a new non-trivial fixed point.  The associated
critical properties and exponents are briefly discussed
in Appendix B.  Remarkably, as we also show in
Appendix B, this relativistic fixed point is in fact linearly {\it stable}
to small deviations in the three velocities.  A microscopic model with
different velocities scales into a relativistic form
near criticality.  As further discussed in Appendix B,
this model is closely related to a remarkable
model\cite{Seiberg}\ which has has a non-trivial
fixed point in $D=2+1$ with an exact
($N=1$) space-time supersymmetry.\cite{Wess}\
Supersymmetry is powerful enough to determine several {\it exact}
critical exponents in $D=3$,\cite{Seiberg}\ which serve as a useful check
on our $\epsilon$ expansion results.

It is finally worth mentioning the effects of the coupling
to the nodons, on the {\it location} of the antiferromagnetic
ordering transition.  With $r_N$ positive one can safely integrate
out the nodons.  This leads to a supression of $r_N$,
with a ``renormalized value", $\tilde{r}_N$, given to leading order by,
\begin{equation}
\tilde{r}_N = r_N - (c\Lambda/v_{\scriptscriptstyle\Delta}) g^2 + O(g^4)  .
\label{r-ren}
\end{equation}
Here $\Lambda$ is the high momentum cutoff on the nodons, and $c$ a cut-off
dependent constant of order one.  The coupling to the nodons
thus tends to {\it enhance} the magnetic ordering.
Within mean-field theory, the antiferromagnetic ordering transition
will take place when the ``renormalized" coefficient
$\tilde{r}_N = 0$, as indicated in Fig.~3a.

\subsubsection{Effects of Particle-hole asymmetry}

Staying at half-filling, we next consider the effects of particle-hole
asymmetries, generated for example by a second neighbor electron
hopping term in a microscopic square lattice model.  As discussed in
Section II, a particle-hole asymmetry generates an additional term in
the nodon Lagrangian of the form,
\begin{equation}
{\cal L}_\lambda = \lambda \psi^\dagger_j \tau^z \psi_j  .
\end{equation}
In the absence of coupling to the N\'eel field, this simply
causes a momentum space shift in the {\it positions} of
the nodes, by an amount $\delta q = \lambda/v_F$.
As we shall see, this has a profound effect when the nodons
are coupled to the N\'eel order parameter, since
such a shift destroys the ``nesting" of the nodons.
Indeed, this leads to two {\it additional} phases at half-filling,
as shown in the phase diagram for $\lambda \ne 0$
in Fig.~3b.

Since both additional phases are antiferromagnetically ordered,
we once again integrate out $\Phi$ and $a_\mu$ (valid for 
non-zero $r_\Phi$), and put $\langle {\bf N} \rangle = N_0 \hat{\bf{y}}$ into
the effective Lagrangian to arrive at a quadratic
nodon Lagrangian of the form ${\cal L}_{nodon} + {\cal L}_\lambda$,
with ${\cal L}_{nodon}$ given in Eq.~\ref{Lnodon}.
Once again a Bogoliubov transformation diagonalizes
the quadratic form, and with non-zero $\lambda$
the energy eigenvalues (in the $j=1$ nodon sector) satisfy,
$E^4 - 2AE^2 + B =0$ with,
\begin{equation}
A = (gN_0)^2 + (v_F q_x)^2 + (v_{\scriptscriptstyle\Delta} q_y)^2 + \lambda^2 ,
\end{equation}
and
\begin{equation}
B = [A - 2\lambda^2] + (2\lambda v_{\scriptscriptstyle\Delta} q_y)^2   .
\end{equation}
If $gN_0 > \lambda$, there is no solution at $E=0$,
so that there is a gap in the nodon spectrum.
The resulting phases -  antiferromagnet
for $r_\Phi < 0$ and co-existing gapped d-wave
superconductivity with antiferomagnetism for $r_\Phi >0$ - 
occur for large negative 
$r_N$, where $N_0$ is large (given by
$N_0 = \sqrt{(-r_N)/2u_N}$ within mean-field theory).
But for smaller $|r_N|$, when $gN_0 < \lambda$,
zero energy solutions {\it do} exist, and there
are {\it gapless} nodon states present!
This leads to the two new phases present in Fig.~3b.

Specifically, for $r_\Phi$ negative, the new phase exhibits
gapless nodon excitations co-existing with long-range antiferromagnetic
order.  The nodons are now incommensurate
with the magnetic order (at $\vec{\pi}$), since the zero
energy nodon state for $\lambda > gN_0$ occurs
at a shifted wavevector: 
\begin{equation}
q_x = v_F^{-1} \sqrt{\lambda^2 - (gN_0)^2} \ne 0 ,
\end{equation}
and $q_y=0$.  This interesting new phase,
which we denote as
AF/NL, exhibits gapless incommensurate
magnetic fluctuations co-existing
with the $\vec{\pi}$ magnons of the N\'eel state. 

For positive $r_\Phi$, where d-wave superconductivity
is present, the gapless nodon excitations 
are simply d-wave quasiparticles, incommensurate
with the N\'eel order.  In the new phase, the
incommensurate gapless d-wave superconductor co-exists
with antiferromagnetic order, as depicted in Fig.~3b

It is worth emphasizing that with
particle-hole asymmetry present (ie. non-zero $\lambda$)
the region of antiferromagnetic order in the phase
diagram is {\it diminished}, relative to
the case with $\lambda = 0$, as depicted in Fig.~3b.
This occurs because incommensurate nodons
are less
effective at renormalizing $r_N$.
Specifically, upon integrating out the nodons with non-zero
$\lambda$ one finds the same form as in Eq.~\ref{r-ren},
but with $c \rightarrow cF(\lambda/v_F \Lambda)$,
where $F(X)$ is a monotonically decreasing function of $X$
with $F(0)=1$.  This leads to a downward shift
in the horizontal phase boundary in Fig.~3b
by an amount, 
\begin{equation}
\delta \tilde{r}_N = [1 - F(\lambda/v_F \Lambda)] 
(c\Lambda/v_{\scriptscriptstyle\Delta}) g^2 .
\end{equation}
Physically, particle-hole asymmetries, such as a second
neighbor hopping term, tend to frustrate and weaken
the antiferromagnetism
at $\vec{\pi}$.  With non-zero $\lambda$
the nodes are shifted off commensurablilty,
and the magnetism is indeed weakened.  This effect
leads to a natural mechanism for the destruction of
antiferromagnetism upon doping, as we describe
in the next section.

Finally, we briefly discuss the nature of the phase transitions
between the six phases present at half-filling with
particle-hole asymmetry.  
Arguments as in the previous subsection,
suggest that the vertical phase boundary
separating the three superconducting from
the three non-superconducting phases (at $r_\Phi =0)$
should, as before, be in the universality class of the
classical $3d$-XY model.   
Since the nodons are incommensurate
at the upper horizontal phase boundary where
antiferromagnetism first appears, they will
decouple from the critical magnetic fluctutations.
Both of these two transitions
(for positive and negative $r_\Phi$) should thus
be in the universality class of the classical
$3d$ Heisenburg model.  At the lower horizontal phase boundaries,
the gapless nodons disappear.  The critical properties are correctly
described by the quadratic nodon Lagrangian, considered above.
In particular, for $\lambda > gN_0$, one can linearize 
for small momentum around the shifted zero energy nodes
by putting $\delta q_x = q_x - v_F^{-1} \sqrt{{\lambda^2}-{(gN_0)^2}}$,
which gives,
\begin{equation}
E = \pm {\left[ \left(1-{\left(\frac{{gN_0}}{\lambda}\right)^2}\right){(v_F
\delta q_x)^2}
+ {v_{\scriptscriptstyle\Delta} q_y^2} \right]^{\frac{1}{2}}}  .
\label{transvelocity}
\end{equation}
As the transition is approached, the velocity
along the $x$-direction -- i.e. perpendicular to the
Fermi surface -- vanishes and the nodons become
quasi-1D.

\subsection{Doping the Antiferromagnet}

We are now in a position to extend our understanding of doping to 
include AF order at half-filling, in accord with experimental 
observations.  In all high-$T_{c}$ materials, doping is actually 
achieved by chemical substitution/depletion of atoms between the 
$CuO_{2}$ layers.  While it is generally believed that this process 
transfers charge to the $CuO_{2}$ planes, this charge transfer is not 
necessarily proportional to the {\sl chemical} doping, defined as the 
fraction of atoms substituted or removed.  To simplify this 
discussion, however, we shall assume that chemical doping indeed 
corresponds to hole doping, and consider the phase diagram as a 
function of hole concentration $x$.  A further assumption concerns the 
degree of particle/hole asymmetry.  Since the composition is changing 
with doping, the parameter $\lambda$ should in general be a function 
of $x$ (in fact, we expect the asymmetry to increase with $x$).  
To simplify the discussion, we shall further assume that any 
explicit dependence of $\lambda$ on $x$ is weak, and therefore treat the 
effects of doping solely through the chemical potential $\mu$.

At half-filling, the system sustains long-range AF order.  In 
general, we expect a non-zero particle/hole asymmetry, so that this could 
correspond to either the AF or AF/NL state, the latter occuring if $\lambda$ 
is sufficiently large.  We do not believe current experiments 
distinguish the two alternative phases in undoped cuprates.   Since 
by assumption we are varying only $\mu$, and $\mu$  
couples {\sl indirectly} to magnetism via the corresponding dual ``flux'' 
in Eq.~\ref{Lgauge}, the AF order and nodons are effected only once 
charge is added to the system.
What is the effect of such charge doping on the AF?  From 
Eqs.~\ref{Llambda},\ref{Lgauge}, we see that the added charges act 
simply to increase the {\sl effective} particle/hole asymmetry of the 
nodons.  In particular,
\begin{equation}
  \lambda_{eff} = \lambda + {x \over {4\kappa_{0}a_{0}^{2}}}.
  \label{leff}
\end{equation}
Eq.~\ref{leff}\ is an extremely useful result.  Using it, we can 
simply trade the doping $x$ for an effective particle/hole asymmetry 
to determine the fate of the system from the phase diagrams at 
half-filling, Fig.~3.

\begin{figure}
\psfig{figure=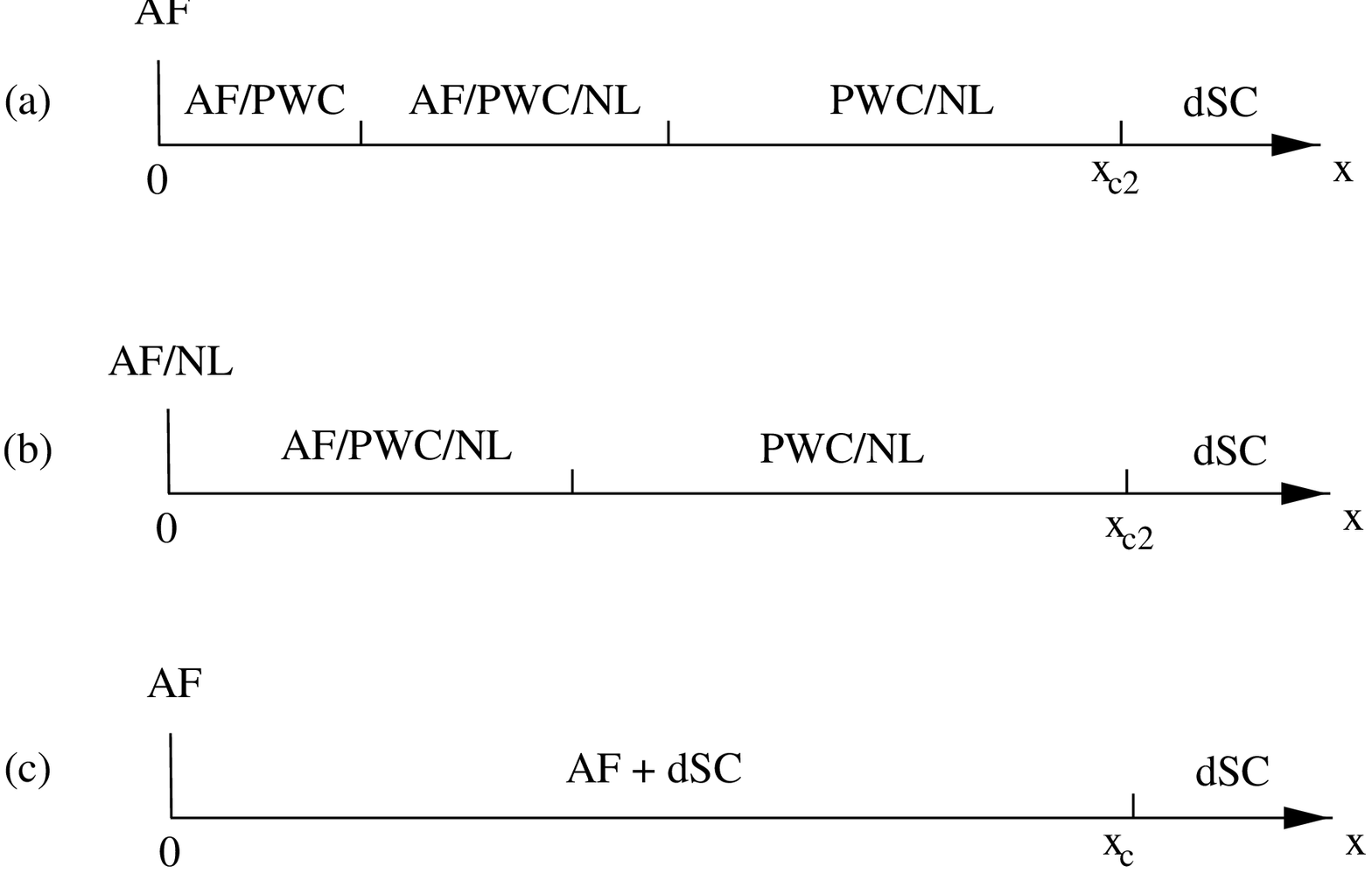,height=1.8in,angle=-90}
\vskip 0.5cm
{Fig.~4: Possible phase diagrams as a function
of doping for the type II ({\bf a} and {\bf b}) and type I ({\bf c})
scenarios.  We tentatively identify the PWC/NL phase, which exhibits neither magnetism or
superconductivity, with the pseudogap state of the underdoped
high-T$_c$ materials.} 
\end{figure}

Consider then first the type II doping scenario.  The charge behavior
is similar to that obtained when doping the NL, Sec.~\ref{sec:NL}.
Upon increasing $\mu$ from zero, the dual ``flux'' is first expelled
from the sample, and the system remains undoped.  Charge first enters
above the dual ``lower critical field'', $\mu>\mu_{c1}$, forming a
Paired Wigner Crystal (PWC) with density $x(\mu)$ due to long-range
Coulomb interactions.  Since $x$ is small at this point, the crystal
coexists with the AF, so the actual phase for small $x$ is an AF/PWC.
As $x$ increases, so does $\lambda_{eff}$, unbinding the nodons into
the AF/PWC/NL.  This can be understood
from the evolution of the phase diagrams at half-filling
as a function of $\lambda$, as shown in Fig.~3.
As $x$ increases further, the NL and AF become
increasingly incommensurate, and the energy gain from their coupling
is eventually reduced sufficiently to destroy the AF order in a
transition to a PWC/NL phase - again a feature
of the phase diagrams at half-filling.  Finally, when $x \geq x_{c2}$, the
upper critical field is reached and the crystal melts into the dSC
phase.  This progression is shown schematically in Fig.~4a.  An
alternate type II doping scenario, shown in Fig.~4b, is that the
system is an AF/NL at half-filling, in which case the phase diagram is
unchanged except for the absence of the AF/PWC phase.

Another
possibility is type I doping.  Because this involves a strong first-order
transition in the absence of Coulomb interactions, the mixed
(micro-phase separated) state could occur as a coexistence between a number
of different phases.  The simplest phase diagram includes only
coexistence between the AF and pure dSC as in Fig.~4c.  Because the
physics of the mixed state is highly non-universal, we do not discuss
it further here.

\section{Discussion}

The main result of this paper is the Lagrangian,
Eqs.~\ref{eq:lagrangian}-\ref{eq:Neel}, which describes the Nodal
Liquid phase, its interaction with external electromagnetic fields,
and transitions between it and the antiferromagnetic (AF) and
superconducting (dSC) phases. This Lagrangian follows directly from
disordering the d-wave superconductor.  It implies that, in the
underdoped region, low-energy fermionic degrees of freedom are located
solely at four isolated (Dirac) points in the Brillouin zone -- a
hypothesis which is strongly supported by ARPES \cite{Loeser,Ding},
NMR \cite{Warren,Takigawa}, optical conductivity \cite{Homes,Puchkov},
and other experiments\cite{Maple}.  The main consequences of
Eqs.~\ref{eq:lagrangian}-\ref{eq:Neel}\ are (1) the prediction of a
new zero-temperature phase, the Nodal Liquid, which interpolates
between the AF and dSC phases; (2) a quantitative description of
charge and spin dynamics in this phase; (3) specific predictions for
the critical behavior at the AF and dSC ordering transitions; and,
above all, (4) a coherent {\it weak-coupling} framework -- with the
Nodal Liquid as its foundation -- for understanding the underdoped
side of the high-$T_c$ phase diagram. We tentatively identify the
coexisting Paired Wigner Crystal/Nodal Liquid (PWC/NL) phase as the
pseudo-gap state in a hypothetical disorder-free underdoped cuprate.
In the real materials, however, disorder will play a role, as we
briefly discuss below.

Our description of this part of the phase diagram enjoys kinship with
several other approaches. The Nodal Liquid phase is reminiscent of the
$\pi$-flux state\cite{Affleck}\ and the $SU(2)$ MFT staggered-flux
state,\cite{Wen}\ and is the d-wave analog of the
short-range\cite{Kivelson}\ resonating valence bond spin-liquid
state.\cite{Anderson}\ These states also involve neutral Dirac
fermions interacting with a gauge field, but the coupling to
electromagnetic fields, the coupling between the fermions and the
gauge fields, the bosonic charged degrees of freedom, and the
evolution with doping are all rather different in the Nodal Liquid.
Several authors\cite{Rice,White}\ have conjectured that the lightly-doped 3-leg
Hubbard ladder might serve as a paradigm for the underdoped cuprates
and Furukawa and Rice\cite{Furukawa}\ have tried to substantiate these claims with
weak-coupling RG calculations on partially nested Fermi liquids.  The
Nodal Liquid concretely realizes the attractive features of this
proposal.  The basic idea of bringing AF and dSC under the same
rubric, which is expressed in (\ref{eq:lagrangian}), is the central
theme of Zhang's $SO(5)$ theory.\cite{Zhang}\  However, there is a
direct transition from AF to dSC in the $SO(5)$ theory, whereas the
Nodal Liquid intervenes in our theory. There is a further important
distinction, namely, that our theory focuses on the zero-temperature
quantum phase transitions of the high-$T_c$ materials.  This is one
reason why our theory accords primary importance to the low-energy
{\sl fermionic} degrees of freedom. Finally, our prediction of phase
separation at the dSC transition
in the type I scenario as well as our
interpretation of $T^*$ echoes the ideas of
Emery and Kivelson.\cite{EK}\

There are a number of important issues which
we have not addressed in this paper.
By restricting our attention to the region underneath the
dashed line in Fig.~1, we
have skirted one of the most controversial
questions in this field: what mechanism drives
pair formation at this scale? Presumably this physics
must be understood for progress to be made on the part
of the phase diagram above and to the right of the dashed line.
This would require an investigation complementary
(but perhaps orthogonal in spirit) to ours.
Also, our discussion of transport was necessarily
incomplete because finite-temperature
transport can be particularly subtle
(for recent examples of this, see Ref.~\onlinecite{Damle})
and the transport properties of the
Nodal Liquid deserve a thorough exposition of their own,
which we defer.  Moreover, a pure sample would {\sl melt} at a finite
temperature phase transition, although this transition would be
rounded by impurities.  The melting temperature is expected to
vanish upon approaching either zero doping or the PWC/NL to dSC
quantum phase transition, and therefore has maximum at some
intermediate $x$ in the underdoped regime.

The effects of disorder are also quite subtle, and warrant a
full and separate treatment.  Nevertheless, a few comments
are germane to this discussion.  The first, and most basic, is that
disorder plays a significant role in the physics of the cuprates. Even
in $Y{Ba_2}{Cu_3}{O_{7-\delta}}$, which is believed to be cleaner
than, say, ${La_{2-x}}{Sr_x}Cu{O_4}$, doping cannot help but introduce
disorder.  According to standard arguments,\cite{Imry}\ first-order
phase transitions will be driven second-order by arbitrarily weak
disorder in two dimensions.  In particular, we expect the
PWC/NL$\rightarrow$dSC transition to be second order with impurities
present.  Moreover, based on the irrelevance (in the technical sense)
of the coupling between the nodons and the superconducting phase in
the clean case, we suspect this transition may be in the same
universality class as the superconductor--insulator transition.  This
could explain the experiments of Fukuzumi, {\it et al.}  on
$Y{Ba_2}{Cu_{3-y}}{Zn_y}{O_{7-\delta}}$.\cite{Fukuzumi}\
The disorder will also have
an effect on the phases themselves. For instance, transport and spin
dynamics in the Nodal Liquid will be influenced by disorder.  Finally,
we note that disorder will transform the PWC into a Bose Glass (BG).
One consequence would be power-law suppression (rather than a hard
gap) of the low frequency {\it electron} spectral function at the
nodes in the BG/NL phase because the BG is compressible.\cite{Fisher89}\  Having
adopted a panoramic view in the preceeding section, we can afford, in
closing, to narrow our focus to the BG/NL (and the PWC/NL from which
it descends) because it is our candidate for the $T=0$ pseudo-gap
phase: a phase without a Fermi surface or long-range order but
possessing low-energy fermionic excitations centered about four points
in the Brillouin zone.

\acknowledgements We are extremely grateful to Doug Scalapino for
numerous helpful conversations.  MPAF would like to thank Steve Girvin
for illuminating discussions on d-wave quasiparticles.  This work has
been supported by the National Science Foundation under grants No.
PHY94-07194, DMR94-00142 and DMR95-28578.

\appendix

\section{Microscopic Approach}

In this appendix, we describe techniques to derive the effective field
theory from some specific microscopic models.  As our purpose
is primarily phenomenological, we will consider one of the simplest
models which develops antiferromagnetism and d-wave superconductivity.
This is a square lattice extended Hubbard model with nearest-neighbor
hopping $t$, on-site electron-electron repulsion $U$ and nearest
neighbor attraction $V$ (we emphasize that this model
is not realistic, but is chosen for illustrative purposes), i.e.
\begin{eqnarray}
  H[c^\dagger,c] & = & -t \sum_{\langle \vec{x}\vec{x}'\rangle}
  \left[c_\alpha^\dagger(\vec{x})
    c_\alpha^{\vphantom\dagger}(\vec{x}') + {\rm h.c.}\right] 
  +\tilde\mu \sum_{\vec{x}} n(\vec{x}) 
     \nonumber \\
  & & + U \left[ 
      n(\vec{x})\right]^2 - V\sum_{\langle
    \vec{x}\vec{x}'\rangle} n(\vec{x})n(\vec{x}').
\end{eqnarray}
Here $n(\vec{x}) = c_\alpha^\dagger(\vec{x})
c_\alpha^{\vphantom\dagger}(\vec{x})$, and $\tilde\mu$ is the
microscopic chemical potential, and we have neglected to include terms
(e.g. a second-neighbor hopping $t'$) which break particle/hole
symmetry at half-filling.  Also, for simplicity we measure
distances in units of the lattice spacing.  The tendency toward AF and
dSC states can brought out by using the identities $[n(\vec{x})]^2 =
-4/3 [{\bf S}(\vec{x})]^2 + 2n(\vec{x})$ and $n(\vec{x})n(\vec{x}') =
c_\alpha^\dagger(\vec{x})c_\beta^\dagger(\vec{x}')
c_\beta^{\vphantom\dagger}(\vec{x}')
c_\alpha^{\vphantom\dagger}(\vec{x})$, for $\vec{x} \neq \vec{x}'$.
The Hamiltonian can be rewritten as
\begin{eqnarray}
  H[c^\dagger,c] & = & -t \sum_{\langle \vec{x}\vec{x}'\rangle}
  \left[c_\alpha^\dagger(\vec{x})
    c_\alpha^{\vphantom\dagger}(\vec{x}') + {\rm h.c.}\right]
  + \mu \sum_{\vec{x}} n(\vec{x}) 
     \nonumber \\
  & & \hspace{-0.5in} - {4 \over 3}U \sum_{\vec{x}} |{\bf S}(\vec{x})|^2 - V\sum_{\langle
    \vec{x}\vec{x}'\rangle}
  c_\alpha^\dagger(\vec{x})c_\beta^\dagger(\vec{x}') 
  c_\beta^{\vphantom\dagger}(\vec{x}') 
  c_\alpha^{\vphantom\dagger}(\vec{x}) ,
  \label{convenient}
\end{eqnarray}
where we defined a shifted chemical potential $\mu$, and have neglected 
an unimportant constant.  As usual, the lattice spin operator is
defined by ${\bf S}(\vec{x}) = {1 \over 2}
c_\alpha^\dagger(\vec{x})\bbox{\sigma}_{\alpha\beta}
c_\beta^{\vphantom\dagger}(\vec{x})$.  The angular brackets $\langle
\vec{x}\vec{x}'\rangle$ under the two sums indicate sums over
all nearest-neighbor pairs of sites.

To derive an effective field theory, it is
convenient to use an {\sl imaginary time} path integral formulation.
In this case one studies the partition function $Z= {\rm Tr}
e^{-H/T}$, where $T$ is the temperature.  It can be represented using
Grassman coherent states as
\begin{equation}
  Z = \int [d\overline{c}][dc] e^{-S},
\end{equation}
where the Euclidean action is
\begin{equation}
  S = \int \! d\tau \left\{ \sum_{\vec{x}}\overline{c}_\alpha(\vec{x})
    \partial_\tau c_\alpha(\vec{x}) 
    + H[\overline{c},c] \right\}.
\end{equation}
We consider here only $T=0$, for which the $\tau$ integration domain
is infinite.  The possibility of dSC and AF phases can be entertained
by decoupling the above action using Hubbard-Stratonovich
transformations.  One finds that
\begin{equation}
  Z = \int [d\overline{c}][dc][d{\bf M}][d\Delta][d\overline\Delta]
  e^{-S_1}, 
\end{equation}
with $S_1 = \int \! d\tau [\sum_{\vec{x}}\overline{c}_\alpha(\vec{x})
    \partial_\tau c_\alpha(\vec{x}) +H_{eff}]$.  The effective
    Hamiltonian can be decomposed into $H_{eff} = H_{qp} +
    H_M + H_{\scriptscriptstyle\Delta}$, with $H_{qp}=H_0+H_{int}$, and  
\begin{eqnarray}
  H_0 & = & -t \sum_{\langle \vec{x}\vec{x}'\rangle}
  \left[c_\alpha^\dagger(\vec{x})
    c_\alpha^{\vphantom\dagger}(\vec{x}') + {\rm h.c.}\right] +
  \mu\sum_{\vec{x}}  n(\vec{x}) \label{Heff0}
   \\
  H_{int} & =  &  - \sum_{\vec{x}} {\bf M}(\vec{x})\cdot {\bf
    S}(\vec{x}) \nonumber \\
  & & + \sum_{\langle \vec{x}\vec{x}'\rangle} \left[
    \overline{\Delta}^{\alpha\beta}_{\vec{x}\vec{x}'}
    c_\beta(\vec{x}')c_\alpha(\vec{x}) + \Delta^{\alpha\beta}_{\vec{x}\vec{x}'}
    \overline{c}_\alpha(\vec{x})\overline{c}_\beta(\vec{x})\right],
  \label{Heffint}  \\
  H_M & = & {3 \over {8U}}\sum_{\vec{x}} |M(\vec{x})|^2, \\
  H_{\scriptscriptstyle\Delta} & = & {1 \over V} \sum_{\langle
    \vec{x}\vec{x}'\rangle}
  \overline{\Delta}^{\alpha\beta}_{\vec{x}\vec{x}'}
  \Delta^{\alpha\beta}_{\vec{x}\vec{x}'} . \label{Heffdelta}
\end{eqnarray}

Eqs.~\ref{Heff0}-\ref{Heffdelta}\ form a basis for studying the
original extended Hubbard model.  Following the philosophy of
Sec.~\ref{sec:duality} , we imagine integrating out high-frequency
modes in the functional integral to arrive at an effective field
theory for the low-lying degrees of freedom.  In the process, one will
generate dynamics for the order parameter $\Delta$ and the
magnetization ${\bf M}$.  For the most part, symmetry considerations
require the corresponding Lagrangians to take the forms given in
Sec.~\ref{sec:duality}\ and Sec.~\ref{sec:Neel}, so we choose not to
complicate the presentation by explicitly performing these
integrations (e.g.  diagrammatically).  

One subtle point in the analysis of Sec.~\ref{sec:duality}, however,
does warrant a more careful treatment.  This is the coupling of the
nodons to the superconducting phase-gradient, from which follows the
expressions for the quasiparticle current, Eqs.~\ref{eq:J0}-\ref{eq:Jj}.  In
Sec.~\ref{sec:duality}, we derive these using the ``symmetric''
prescription of Eq.~\ref{dgauge}.  We now show that the
currents are indeed obtained correctly using this prescription.

We first specialize to the case of singlet pairing,
$\Delta^{\alpha\beta}_{\vec{x}\vec{x}'} = \Delta(\vec{x},\vec{x}')
(\delta_{\alpha\uparrow}\delta_{\beta\downarrow} -
\delta_{\alpha\downarrow}\delta_{\beta\uparrow})$.  Since $\Delta$
lives on the bonds, it is convenient to associate two such fields with 
each site in the square lattice, i.e.
\begin{eqnarray}
  \Delta_1(\vec{x}) & \equiv & \Delta(\vec{x},\vec{x}+\hat{e}_1),
  \label{pv1} \\
  \Delta_2(\vec{x}) & \equiv & \Delta(\vec{x},\vec{x}+\hat{e}_2),\label{pv2}
\end{eqnarray}
where $\hat{e}_1, \hat{e}_2$ are unit vectors along the $a$ and $b$
axes of the square lattice, respectively.  Note that at this point, we
have specified no particular relation between $\Delta_1$ and
$\Delta_2$, so that the model has the potential both for d-wave and
s-wave pairing.  To distinguish them, we must consider the form of the
effective action for $\Delta,\overline{\Delta}$ generated upon
integrating out the high-energy modes.  By symmetry, the simplest
local allowed additional term on the lattice is a sum of
$U(1)$-invariant two-bond products around each lattice site, which can 
be written as $S_2 = \int\! d\tau\, H_2$, with 
\begin{eqnarray}
  H_2 &  = & {\tilde\gamma \over 8}
  \sum_{\vec{x}} \bigg\{ 
  \overline{\Delta}_1(\vec{x})\Delta_2(\vec{x}) +
  \overline{\Delta}_2(\vec{x})\Delta_1(\vec{x}-\hat{e}_1) \nonumber \\
  & & \hspace{-0.2in} +
  \overline{\Delta}_1(\vec{x}-\hat{e}_1)\Delta_2(\vec{x}-\hat{e}_2) +
  \overline{\Delta}_2(\vec{x}-\hat{e}_2)\Delta_1(\vec{x}) + {\rm
    c.c.}\bigg\}. \nonumber \\
  & & \label{eq:around}
\end{eqnarray}
Of course, the actual quadratic $\overline{\Delta}\Delta$ interaction
terms generated from the high-energy degrees of freedom will be much
more complex.  However, since the general form of the {\sl
  long-wavelength} effective action is dictated by symmetry, this
example suffices for illustrative purposes.  Writing $\Delta_j =
\Delta_0 e^{i\phi_j}$, Eq.~\ref{eq:around}\ becomes
\begin{eqnarray}
  H_2 &  = & {\gamma \over 4} 
  \sum_{\vec{x}} \bigg\{ \cos(\phi_1(\vec{x})\!-\!\phi_2(\vec{x})) +
  \cos(\phi_2(\vec{x})\!-\!\phi_1(\vec{x}\!-\!\hat{e}_1)) \nonumber \\
  & & \hspace{-0.2in}+
  \cos(\phi_1(\vec{x}\!-\!\hat{e}_1)\!-\!\phi_2(\vec{x}\!-\!\hat{e}_2)) + 
  \cos(\phi_2(\vec{x}\!-\!\hat{e}_2) \!-\!\phi_1(\vec{x}))\bigg\},
  \nonumber \\
  & &   \label{Sdelta}
\end{eqnarray}
with $\gamma = (\Delta_0)^2\tilde\gamma$.  We assume $\gamma>0$, in
which case this interaction favors a relative phase difference of
$\phi_1-\phi_2 = \pi$, i.e. $d_{x^2-y^2}$ order.

We now turn to the fluctuations around the uniform dSC state.  To do
so, we let $\phi_1 = \varphi$, $\phi_2 = \varphi + \theta + \pi$.  The 
phase $\varphi$ is precisely the order parameter phase introduced in
Sec.~\ref{sec:duality}.  The other variable $\theta$ represents
another branch of {\sl massive} fluctuations around the d-wave state.
We can thus assume $\theta \ll 1$, and that $\varphi$ is slowly
varying, i.e. $\partial_j\varphi \ll 1$.  Under this assumption, we
can take the continuum limit and replace the positional sum in
Eq.~\ref{Sdelta}\ by an integration.  This gives $H_2 = \int\! d^2x
{\cal H}_2$, with
\begin{equation}
  {\cal H}_2 = \gamma \bigg\{
  {\theta^2 \over 2} + {1 \over 4}(\partial_j\varphi)^2 -
  {1 \over 4}\partial_x\varphi\partial_y\varphi + {\theta \over
    2}(\partial_x-\partial_y)\varphi \bigg\}.
\end{equation}
As expected, the $\theta$ field is massive, and can be integrated
out.  Equivalently, one minimizes ${\cal H}_2$ with respect to
$\theta$.  This process restores isotropy and gives
\begin{equation}
  {\cal H}_2 \rightarrow {\gamma \over 8}(\partial_j\varphi)^2.
\end{equation}
This is just the spatial component of the superfluid Lagrangian,
Eq.~\ref{applag}, with $\gamma = v_c^2 \kappa_0$.  The corresponding time
component can be obtained similarly.

Finally, consider the coupling of of the phase $\varphi$ to the
fermionic quasi-particle operators.  To study this, we take for
simplicity $\mu = {\bf M} = 0$.  Using the definitions in
Eqs.~\ref{pv1}-\ref{pv2}, the coupling term in Eq.~\ref{Heffint}\ 
can be rewritten
\begin{eqnarray}
  H_{int} & = & \sum_{j,\vec{x}} \Bigg\{ \Delta_j(\vec{x})
  \left[c^\dagger_\uparrow(\vec{x})
    c^\dagger_\downarrow(\vec{x}+\hat{e}_j) -
    \uparrow\leftrightarrow\downarrow\right] + {\rm h.c.}\Bigg\} ,    
  \label{d12}
\end{eqnarray}
where the sum includes all lattice sites and $j=1,2$.  To ease
comparison with Sec.~\ref{sec:duality}, we have returned now to a
Hamiltonian formalism, replacing $\overline{c}_\alpha$ by
$c_\alpha^\dagger$. We are now in a position to take the continuum
limit.  In this case, it suffices to neglect the massive $\theta$
mode, and simply take $\Delta_1 = -\Delta_2 = \Delta_0 e^{i\varphi}$.
For agreement with Sec.~\ref{sec:duality}, we define
$v_{\scriptscriptstyle\Delta} = 2\sqrt{2}\Delta_0$, or $\Delta_1 =
-\Delta_2 = \tilde\Delta/2\sqrt{2}$.  In addition, we take the
continuum limit of the electron fields, using the decompositions
\begin{eqnarray}
  c^\dagger_\uparrow & \sim & \Psi_{111}^\dagger i^{x+y} \!-\!
  \Psi_{122}^{\vphantom\dagger}(-i)^{x+y}\!+\!\Psi_{211}^\dagger(-i)^{x-y} 
  \!-\!\Psi_{222}^{\vphantom\dagger}i^{x-y}, \nonumber \\
  c^\dagger_\downarrow & \sim & \Psi_{112}^\dagger i^{x+y} \!+\!
  \Psi_{121}^{\vphantom\dagger}(-i)^{x+y}\!+\!\Psi_{212}^\dagger(-i)^{x-y} 
  \!+\!\Psi_{221}^{\vphantom\dagger}i^{x-y},\nonumber
\end{eqnarray}
and the hermitian conjugates of these equations.  Inserting these into
Eq.~\ref{d12}, gradient-expanding the $\Psi$ fields, and rotating $45$ 
degrees to $x-y$ coordinates along the $(\pi,\pi)$ and $(-\pi,\pi)$
directions, one obtains
$H_{int} = \int\!d^2{x} {\cal H}_{int}$, with
\begin{eqnarray}
  {\cal H}_{int} & = & \left[ {\tilde\Delta \over 2} \left(\Psi_1^\dagger \tau^+
    i\partial_y \Psi_1^{\vphantom\dagger} -
    (i\partial_y\Psi_1^\dagger) \tau^+
    \Psi_1^{\vphantom\dagger}\right) + {\rm h.c.}\right] \nonumber \\
& & + (1 \leftrightarrow 2, x\leftrightarrow y).
\end{eqnarray}
This form is identical to the $\tilde\Delta$ term in Eq.~\ref{hpsi}\ when the
order parameter $\tilde\Delta$ is constant, but the symmetric
placement of derivatives is important in the presence of phase
gradients.  In particular, now let $\tilde\Delta =
v_{\scriptscriptstyle\Delta}e^{i\varphi}$ and integrate by parts to
transfer the derivative in the second term from the $\Psi^\dagger$ to
the $\tilde\Delta\Psi$ combination.  Then, using the operator identity
\begin{equation}
  {1 \over 2} \left(e^{i\varphi}i\partial_y + i\partial_y
    e^{i\varphi}\right) = e^{i\varphi/2} i\partial_y e^{i\varphi/2},
\end{equation}
one obtains 
\begin{equation}
  {\cal H}_{int} =   \sum_{s=\pm} \Psi_1^\dagger [
  v_{\scriptscriptstyle\Delta} \tau^s  e^{is\varphi/2}
  (i \partial_y) e^{is\varphi/2} ] 
  \Psi_1 + (1 \leftrightarrow 2; x \leftrightarrow y).
  \label{Hintresult}
\end{equation}
Eq.~\ref{Hintresult}\ is identical to the symmetrized form of the
phase-quasiparticle interaction hypothesized in Eq.~\ref{hqp}.

\section{Renormalization Group Analysis}

In this appendix, we present some
details of the RG calculations for the
transition at half-filling with particle/hole symmetry
between the nodal liquid phase 
and the antiferromagnet.  As discussed in Section IV,
this transition is described by the {\it same}
field theory as the phase transition between the d-wave superconducting
phase and the phase
with co-existing antiferromagent and gapped d-wave
order.
A striking feature of this transition, which we access
perturbatively in an $\epsilon = 4 -(d+1)$ expansion,
is that it is Lorentz-invariant.  Although the model possesses
{\it three} independent velocities, the differences between
these scale to zero at the critical point.

The full Lagrangian is given by,
\begin{eqnarray}
\label{neellagvel}
{\cal L} = & & {\psi^\dagger_{1}}\left(i{\partial_t}-
{v_F}{\tau^z}{i\partial_x}
- {v_{\scriptscriptstyle\Delta}}{\tau^x}{i\partial_y}\right) {\psi_{1}}\cr
&+& {\psi^\dagger_{2}}\left(i{\partial_t}-
{v_{\scriptscriptstyle\Delta}}{\tau^z}{i\partial_x}
- {v_F}{\tau^x}{i\partial_y}\right) {\psi_{2}}\cr
&+& \frac{1}{2} K_\mu |\partial_\mu {\bf N}|^2 - {r_N}{{\bf N}^2} - u {\left({{\bf 
N}^2}\right)^2}\cr
&+& g {\bf N}\cdot [\psi^\dagger \tau^y \bbox{\sigma}
\sigma^y \psi^\dagger + c.c.] 
  ,
\end{eqnarray}
where we have suppressed the spin subscripts $\alpha,\beta$
on the $\bbox{\sigma}$ Pauli matrices, and the particle/hole
subscripts $a,b$ 
on the $\bbox{\tau}$ Pauli matrices.
Here $K_0=K$ and $K_j = -v_s^2 K$, for $j=1,2$.
Notice that this model has three independent velocities,
$v_F, v_{\scriptscriptstyle\Delta}$ and $v_s$.

Now, we can rescale the ${\bf N}$ field to set $K=1$,
and rescale $\vec{x}$ to set $v_s=1$. 
The Lagrangian can then be rewritten as:
\begin{eqnarray}
\label{neellag-vel}
{\cal L} = & & {\psi^\dagger_{j \alpha}}\left(i{\partial_t}-
{\tau^z}{i\partial_x}
- {\tau^x}{i\partial_y}\right) {\psi_{j \alpha}}\cr
&+& \frac{1}{2}{\left({\partial_t}{\bf N}\right)^2}
- \frac{1}{2}{\left({\partial_j}{\bf N}\right)^2}
- {r_N}{{\bf N}^2} - u {\left({{\bf N}^2}\right)^2}\cr
&+& g {\bf N}\cdot [\psi^\dagger \tau^y {\bf \sigma} \sigma^y \psi^\dagger + c.c.] 
\cr
&-& {\psi^\dagger_{1}}\left(
{a_{1}} {\tau^z}{i\partial_x}
+  {a_{2}} {\tau^x}{i\partial_y}\right)
{\psi_{1}}\cr
&-& {\psi^\dagger_{2}}\left(
{a_{2}}{\tau^z}{i\partial_x}
+  {a_{1}} {\tau^x}{i\partial_y}\right) {\psi_{2}}
\end{eqnarray}
where
\begin{eqnarray}
{a_{1}}&=&\frac{v_F}{v_{s}}-1   ,\cr
{a_{2}}&=&\frac{v_{\scriptscriptstyle\Delta}}{v_{s}}-1  .
\end{eqnarray}

Power counting about the Gaussian theory reveals that both
$u$ and $g^2$ are relevant in $D=2+1=3$ space-time dimensions,
but become marginal in $D=4$.
To implement a perturbative RG calculation thus requires
continung the model above $D=3$.
This is a little tricky, due to the Pauli-matrix algebra.
One approach is to dimensionally continue the loop integrals,
but leave the $3D$ Pauli-matrix
algebra unchanged. 
This turns out to be equivalent to introducing into
the Dirac equations an extra
Pauli matrix, $\tau^y$, multiplied by another (3rd)
spatial dimension.  Alternatively,
one could  
replace the two-dimensional Pauli matrices by the
$4-$dimensional $\gamma$-matrices,
appropriate for $4D$ spinors.
This latter procedure would be more
correct if we were truly interested in the vicinity of four dimensions, but
our choice probably makes more sense given that we are eventually concerned
with $\epsilon \rightarrow 1$. In any case, the difference is
fairly trivial: factors of $2$ would be replaced by factors of $4$ in
traces over the $4D$ $\gamma$-matrices. With
our convention, we obtain the following one-loop
flow equations:
\begin{eqnarray}
\label{floweqns}
\frac{du}{dl} &=& \epsilon u - 44 {u^2} - 32 \lambda u + 32 {\lambda^2}\cr
\frac{d\lambda}{dl} &=& \epsilon \lambda - 20 {\lambda^2}\cr
\frac{d{r_N}}{dl} &\equiv& \frac{1}{\nu}\,{r_N} = (2-20u-16\lambda){r_N}\cr
\frac{d{a_{j}}}{dl} &=& - 12\lambda {a_{j}} ,
\end{eqnarray}
where $\lambda={g^2}$. ${\bf N}$ and $\psi$ have anomalous dimensions
$8\lambda$ and $3\lambda$, respectively.  These flow equtions have a
fixed point at $O(\epsilon)$: $\lambda=\epsilon/20$, $u=
\epsilon\sqrt{371}/440$, ${r_N}=0$, ${a_{j}}=0$. This fixed point has
the following interesting features:

\noindent{(1) `Relativistic Invariance.' Since the velocity differences
scale to zero according to (\ref{floweqns}), all
physically measurable quantities are a function
of ${x^2}+{y^2}-{t^2}$ in the units which we have chosen,
or, upon restoring the velocities,
$({x^2}+{y^2})-v_s^2{t^2}$.}

\noindent{(2) An antiferromagnetic correlation length which diverges
as}
\begin{equation}
\label{corrlength}
\xi \sim |{r_N} - {r_N^c}|^{-1/(2-20{u^*}-16{\lambda^*})}
\end{equation}
as the transition is approached.

\noindent{(3) Critical correlation functions with the following
power-law decays}
\begin{eqnarray}
\langle {N_i}(x,t){N_j}(0,0)\rangle
&\sim& \frac{{\delta_{ij}}}{|{{\vec{x}^2}-{t^2}}|^{\frac{1}{2}-16{\lambda^*}}}
\cr
\langle {\psi^\dagger}(x,t)\psi(0,0)\rangle
&\sim& \frac{1}{|{{\vec{x}^2}-{t^2}}|^{1-6{\lambda^*}}}
\end{eqnarray}

As a check on the reliability of the $\epsilon$-expansion,
we consider the following related model\cite{Seiberg}:
\begin{eqnarray}
{\cal L} =& &{\psi^\dagger}\left(i{\partial_t}-
{\tau^z}{i\partial_x}
- {\tau^x}{i\partial_y}\right) {\psi}\cr
&+& \frac{1}{2}{\left|\partial_t\Phi\right|^2} - 
\frac{1}{2}{\left|\partial_j\Phi\right|^2}
- u {\left|\Phi\right|^4} + g [\psi^\dagger {\tau^y}\psi^\dagger + {\rm c.c.}] , 
\end{eqnarray}
where $\psi$ is now a single two-component spinor, and $\Phi$
is a complex field.
Using the same $\epsilon$-expansion procedure, we
find a fixed point at ${\lambda^*}={u^*}=\epsilon/6$;
at this fixed point, the fields $\Phi$ and
$\psi$ have anomalous dimension $\epsilon/6$.
It is a remarkable and fortunate fact that
this model exhibits $N=1$ supersymmetry. As a result,
the existence of a fixed point is guaranteed and the scaling
dimensions of $\psi$ and $\Phi$ can be determined
exactly, in agreement with the
$\epsilon$ expansion.

\end{multicols}

\end{document}